\documentclass[10pt]{bmc_article}

\usepackage{cite} 
\usepackage{url}  
\usepackage{ifthen}  
\usepackage{multicol}   
\usepackage{listings}
\usepackage{comment}
\usepackage{rotating}
\usepackage[utf8]{inputenc} 
\usepackage{amsmath,amssymb,graphics,graphicx} 
\usepackage{color}

\newboolean{publ}



\newenvironment{bmcformat}{\begin{raggedright}\baselineskip20pt\sloppy\setboolean{publ}{false}}{\end{raggedright}\baselineskip20pt\sloppy}

\begin{document}
\begin{bmcformat}


\title{The plasticity of TGF-$\beta$ signaling}
 

\author{Geraldine Celli\`ere$^{1,\dag}$ \and Georgios Fengos$^{1,2, \dag}$ \and Marianne Herv\'e$^{1}$ and Dagmar Iber$^{1,2,*}$}


\address{$^{1}$ Department of Biosystems Science and Engineering, Eidgen\"{o}ssische Technische Hochschule Zurich, Mattenstrasse 26, 4058 Basel, Switzerland \\ $2$ Swiss Institute of Bioinformatics \\ $\dag$ The authors contributed equally to the work. }

\email{DI: dagmar.iber@bsse.ethz.ch
GC: gcellier@student.ethz.ch
GF: georgios.fengos@bsse.ethz.ch
MH: mh.herve@gmail.com}

\maketitle


\newpage

\begin{abstract}
        \paragraph*{Background:} 
        The family of TGF-$\beta$ ligands is large and its members are involved in many different signaling processes. These signaling processes strongly differ in type with TGF-$\beta$ ligands eliciting both sustained or transient responses. Members of the TGF-$\beta$ family can also act as morphogen and cellular responses would then be expected to provide a direct read-out of the extracellular ligand concentration. A number of different models have been proposed to reconcile these different behaviours. We were interested to define the set of minimal modifications that are required to change the type of signal processing in the TGF-$\beta$ signaling network.
       
        \paragraph*{Results:} 
To define the key aspects for signaling plasticity we focused on the core of the TGF-$\beta$ signaling network. With the help of a parameter screen we identified ranges of kinetic parameters and protein concentrations that give rise to transient, sustained, or oscillatory responses to constant stimuli, as well as those parameter ranges that enable a proportional response to time-varying ligand concentrations (as expected in the read-out of morphogens). A combination of a strong negative feedback and fast shuttling to the nucleus biases signaling to a transient rather than a sustained response, while oscillations were obtained if ligand binding to the receptor is weak and the turn-over of the I-Smad is fast. A proportional read-out required inefficient receptor activation in addition to a low affinity of receptor-ligand binding. We find that targeted modification of single parameters suffices to alter the response type. The  intensity of a constant signal (i.e. the ligand concentration), on the other hand, affected only the strength but not the type of the response. 

        \paragraph*{Conclusions:} 

The architecture of the TGF-$\beta$ pathway enables the observed signaling plasticity. The observed range of signaling outputs to TGF-$\beta$ ligand in different cell types and under different conditions can be explained with differences in cellular protein concentrations and with changes in effective rate constants due to cross-talk with other signaling pathways.  It will be interesting to uncover the exact cellular differences as well as the details of the cross-talks in future work.
        
\end{abstract}

\ifthenelse{\boolean{publ}}{\begin{multicols}{2}}{}


\newpage

\section*{Background}

Transforming growth factor beta (TGF-$\beta$) signaling has been implicated as an important regulator of almost all major cell behaviors, including proliferation, differentiation, cell death, and motility \cite{Blobe2000}. Which response is induced or repressed depends on the cell type and context in which the signal is received.  \\[0.3cm]

The complexity of the biological outcomes elicited  by TGF-$\beta$ stands  in stark contrast to the apparent simplicity of the 
signaling cascade. In response to TGF-$\beta$, type 1 (ALKs 1-7 in humans) and type 2 receptors 
(ActR-IIA, ActR-IIB, BMPR- II, AMHR-II and TbR-II, in humans) form complexes and the constitutively active type 2 serine/threonine kinase phosphorylates the type 1 receptor. The activated type 1 receptor transduces the signal into the cell by phosphorylating the regulatory Smads (R-Smad: Smad 2 and 3 in case of the TGF-$\beta$ subfamily, and Smad 1,5 and 8 for the BMP subfamily). Once activated R-Smads form homomeric complexes and heteromeric complexes with the common Smad, Co-Smad (Smad 4) \cite{Shi2003}. Smads continuously shuttle  between nucleus and cytoplasm \cite{Schmierer2008}. TGF-$\beta$ signaling biases Smad localisation to the nucleus  \cite{Pierreux2000} where Smad complexes associate with chromatin and regulate the transcription of hundreds of genes \cite{Kang2003}. Signal termination is achieved through continuous dephosphorylation of the R-Smad (mainly in the nucleus \cite{Schmierer2008}) and induction of inhibitory Smads (I-Smad: Smad 6 for the BMP subfamily, and Smad7 for the TGF$\beta$ subfamily). I-Smads act through diverse mechanisms: by targeting active receptor for proteasomal degradation \cite{Kavsak2000, Ebisawa2001}, inducing receptor dephosphorylation \cite{Randall2002} and competing with R-Smad for the receptor binding site \cite{Hayashi1997}. Rapid shuttling and inactivation enables a continuous sensing of the extracellular ligand concentrations \cite{Schmierer2008}. This is likely to be particular important when members of the TGF-$\beta$ ligand family acts as morphogen and determine cell-fate in a concentration-dependent manner. \\[0.3cm]

Beyond the core components of this signaling pathway many other factors modulate the signal and thereby contribute to the versality of the  response. At the membrane level, the access to receptor is controlled by soluble proteins that 
sequester TGF-$\beta$ ligand (i.e. decorin) \cite{Massague2000b}, and by membrane-bound co-receptors that 
promote binding (i.e. betaglycan) \cite{Esparza2001}. The receptor activity is further regulated by several receptor 
internalization routes \cite{Guglielmo2003}, and by receptor turnover. Intracellularly, many processes require auxiliary proteins (i.e. SARA for the binding of R-Smad to the receptor and Schnurri for the binding of the R-Smad/Co-Smad complex to the DNA binding element) \cite{Shi2003, Pyrowolakis2004}. The restriction of those auxiliary factors to specific cell-types will make the response cell context dependent \cite{Moustakas2001}. Diversity can also be generated by the huge number of different possible combinations of type 1 and type 2 receptors \cite{Shi2003} and the multiple crosstalks of the TGF-$\beta$ signaling cascade with other pathways. One example of regulation by cross-talk is the phosphorylation of R-Smads in the linker region by Ras-activated MAPK \cite{Kretzschmar1999}, calcium-calmodulin-dependent protein kinase II \cite{Wicks2000} or CDKs \cite{Matsuura2004}. Phosphorylation reduces the transcriptional activity of the  R-Smad \cite{Grimm2002}. \\[0.3cm]

Several mathematical models have been developed to gain further insights into the complex TGF-$\beta$-dependent signaling network \cite{Clarke2008}. An early model by Clarke and co-workers (2006) \cite{Clarke2006} focused on the nuclear accumulation of Smad complexes. Their conclusion on the central role of the imbalance between R-Smad phosphorylation and dephosphorylation rates were confirmed by a more detailed model by Schmierer \textit{et al.} (2008) \cite{Schmierer2008}. Experiments suggest that the duration of the response to a ligand stimulation strongly impacts on the cellular response. Thus epithelial cells that elicit sustained nuclear Smad complex accumulation respond to TGF-$\beta$ with cell growth arrest, whereas pancreatic tumor cells that elicit a transient response continue proliferating (while keeping other TGF-$\beta$ induced behaviors) \cite{Nicolas2003}. Much theoretical work therefore focused on how sustained, transient, or switch-like responses could be obtained by adjusting the receptor dynamics, ligand depletion, and the I-Smad dependent negative feedback. Melke \textit{et al.} (2006) \cite{Melke2006} focused on the potential role of I-Smads in generating transient responses while Vilar \textit{et al.} (2006) focused on the receptor dynamics to explain the occurrence of both transient and sustained responses.  Zi \textit{et al.} (2007) \cite{Zi2007} included a simple model of the Smad dynamics and highlighted the importance of the balance between clathrin-dependent endocytosis and non-clathrin mediated endocytosis. All pathway elements were finally brought together by Chung \textit{et al.} (2009) \cite{Chung2009} in a more comprehensive model, used to examine the contradictory roles of TGF-$\beta$ in cancer progression. Lately Zi \textit{et al.} (2011) \cite{Zi2011} published a study that highlights the potential of TGF-$\beta$ ligand depletion in converting short-term graded signaling responses into long-term switch-like responses. Unlike for other pathways oscillations have not yet reported for the TGF-$\beta$ signaling pathway \cite{Novak:2008p1606, Shankaran:2009p26964}. TGF-$\beta$ type ligands are also acting as morphogens, and the response to these appears to be proportional. Recently, Paulsen and co-workers published a study on the impact of synexpression of the feedback inhibitors BAMBI, Smad6, and Smad7 on the read-out of morphogen gradients during embryogenesis \cite{Paulsen:2011gs}.   \\[0.3cm]

While the many published studies explain the different behaviours for the different situations for which they are observed and highlight the many mechanisms that enable the different response types it remains largely unclear how easily the response type can be changed. We wondered how the TGF-$\beta$ signaling pathway accomplishes the flexibility in its responses and which and how many parameters have to be altered for cells to respond differently. To efficiently explore the canonical response we focused on the core signaling architecture, and did not consider the detailed receptor dynamics and cross-talks in the model; they are included indirectly through the parameters that they modulate. We explored the response types and in particular changes in the response type as we explored the parameter values within biologically meaningful ranges. We find that relatively small changes in single parameters can alter the response. Cellular protein concentrations are a particular powerful point of control and this explains how different cell types can show different responses. Importantly we also identify key parameters that affect the response and we can relate these to observed points of cross-talk between signaling pathways. The particular architecture of the TGF-$\beta$ network  thus allows for the  great flexibility in  the response. \\[0.3cm]

\section*{Methods}

\subsection*{The model}
Several models for the TGF-$\beta$ signaling network have been developed that focus on different aspects of the TGF-$\beta$ signaling network, i.e. the receptor dynamics \cite{Vilar2006, Zi2007}, the shuttling between the cytoplasm and the nucleus \cite{Schmierer2008}, and the negative feedback via the I-Smad (Smad7/Dad) \cite{Melke2006}. These different aspects have lately been combined in a model that addresses differences in TGF-$\beta$ signaling between normal and cancerous cells \cite{Chung2009}. The models of the TGF-$\beta$ signaling pathway showed that stimulation could result in either transient and sustained responses dependent on the choice of parameters \cite{Schmierer2008, Vilar2006, Melke2006, Zi2007, Chung2009, Zi2011}. Transient responses could be obtained through complex receptor dynamic \cite{Vilar2006}, the I-Smad-mediated negative feedback \cite{Melke2006, Schmierer2008}, or ligand depletion \cite{Zi2011}. Negative feedbacks can in principle also give rise to oscillatory behaviour. We wondered whether all three qualitative behaviours (sustained, transient, or oscillatory response) could be obtained already with the most simple intracellular feedback mechanism, and how these behaviours would depend on the parameters. Since the more complex interactions (that we ignore) effectively modulate the parameter values in our model an in-depth understanding of the parameter dependencies in the simple model should also enable a better understanding of the complex network interactions that are found in the cell. The different response types can also (trivially)  be obtained by modulating the protein concentrations accordingly. We, however, keep the concentrations of receptors, ligand, R-Smad and Co-Smad constant and thus include these effects only indirectly as changes in the effective binding rates.  \\[0.3cm]

Accordingly, we formulated a detailed model of TGF-$\beta$ signaling that focused on the negative feedback, but did not include any complex receptor dynamics  as these require changes in the receptor and ligand concentrations. Our model describes the dynamics of TGF-$\beta$ ligand (TGF-$\beta$), receptor ($TGF\beta R$), regulatory R-Smads (denoted simply Smad), Co-Smads, I-Smads, their complexes as well as the expression intermediates of the I-Smad. Importantly, we include two compartments, the nucleus and the cytoplasm, and the Smad and Co-Smad complexes can shuttle between the two compartments as first described in  \cite{Schmierer2008}. The regulatory interactions are summarized in Fig.~\ref{Fig1} (a SBML file is provided in Additional file 3). Thus the ligand TGF-$\beta$ reversibly binds to the TGF-$\beta$ receptor (reactions 1 and 2 in Fig.~\ref{Fig1}), which is then phosphorylated to become fully active (3 and 4). The active receptor induces phosphorylation of R-Smad (7), which in turn can reversibly dimerize or form a complex with Co-Smad (10 and 11). Those two reactions can take place either in the cytoplasm or in the nucleus and the five species Smad, phosphorylated Smad, Co-Smad, homodimers and heterodimers can shuttle from the cytoplasm to the nucleus and back (8, 9 and 12). Nuclear Smad/Co-Smadf complexes act as transcription factors and trigger the transcription of I-Smad mRNA in the nucleus (14 and 15). The I-Smad mRNA then shuttles to the cytoplasm (16), where it can be degraded (17) or translated into I-Smad (18). I-Smad mediates a negative feedback by sequestering the active receptor (5 and 6) and can be degraded (19). The response to a stimulus by TGF-$\beta$ ligand is a change in the transcriptional activity, monitored as the nuclear concentration of Smad/Co-Smad complexes. \\[0.3cm]

We translated those interactions into sets of ODEs using the law of mass action where appropriate. To reduce the complexity of the model we also employed Hill functions to describe the regulation by cooperative interactions. To efficiently  investigate the impact of changes in total concentration of receptors, R-Smad, and Co-Smad we used a total concentration rather than production and degradation rates for these species. \\[0.3cm]

To respond to TGF-$\beta$ cells must be able to detect changes in the ligand concentration and convert the differences into different transcriptional responses. Transcriptional activity is determined by the concentration of transcription factors in the nucleus. We therefore monitor the nuclear concentration of R-Smad/Co-Smad complexes as a measure of transcriptional activity, in response to a change in the extracellular TGF-$\beta$ concentration. \\[0.3cm]

\subsection*{Parameter screening and simulations}

We are interested in the signaling capacity of the TGF-$\beta$ pathway within its physiological limits. These physiological limits are set by the plausible range that the parameter values can take. We established a likely range for each parameter value based on available data and estimates (Additional file 1, Table S1 and Table S2). While previous measurements and estimates are necessarily of limited accuracy and differences are likely to exist between different cells and different cell types \cite{Schmierer2008, Vilar2006, Melke2006, Zi2007,Chung2009} we expect that basing ourselves on the available data will not too much distort the ranges that we screen. Most parameters were varied over 3 or 4 orders of magnitude, centered around the mean of values found in the literature. Since there are no good estimates for the I-Smad expression rates k14 and k15 were varied over 5 orders of magnitude. The rates of phosphorylation and dephosphorylation (k7 and k13) were varied only over two orders of magnitude because a large fraction of the simulations failed when these rate constants were varied over a wider range. To avoid a bias to the few parameter sets that do not lead to extreme dynamics we had to constrain these two parameters to only vary over two orders of magnitude. To determine the possible range of pathway responses to a defined stimulus, we carried out 10$^6$ independent simulations with parameter values randomly picked from a uniform logarithmic distribution of parameter values within the set ranges (as discussed in Geier \textit{et al.} \cite{Geier2011}) and compared the predicted nuclear concentration of R-Smad/Co-Smad complexes in response to the ligand stimulus. In a first step, we let the system equilibrate for 1hour with almost no ligand (concentration of $10^{-6}$pM to avoid failure of the solver) and initial cellular concentrations $TGF\beta R=1nM$, Smad$=60nM$ and Co-Smad$=100nM$. We then used the steady-state value of the first step and solved the simulations for 10hours with a constant ligand concentration of 200pM. Using MATLAB's ode15s routine the 10$^6$ simulations took in total approximately 140 hours of CPU time. \\[0.3cm]

\subsection*{Criteria to define the different TGF-$\beta$ signaling responses}
 
In response to ligand exposure we observed five different qualitative responses, i.e. unresponsive, sustained, transient, dampened oscillatory or sustained oscillatory responses (Fig.~\ref{Fig2}). Additional file 2, Fig.~S1, S2, S3, and S4 show the evolution of the concentration of each species over time in a representative transient and a representative sustained response (Additional file 1, Table S3). To define the parameter dependency of the different response types we made the following definitions: We speak of unresponsiveness if the concentration of nuclear R-Smad/Co-Smad complexes remains below a chosen threshold $\theta$ within ten hours of stimulation. Accordingly we speak of responsiveness if the concentration exceeds the threshold concentration $\theta$, and here we distinguished four distinct behaviours, inspired by the work of Ma \textit{et al.}\cite{Ma2009} and based on the subsequent dynamics:  
 
\begin{enumerate}
\item \textbf{Sustained response:} After the initial peak the response must retain at least 90$\%$ of its maximal value (called Opeak). To exclude slowly increasing responses we further require that 90\% of the peak value Opeak is reached within less  than 7200 s (2 hours).
\item \textbf{Transient response:} After the initial peak the response must drop to levels lower than 10$\%$ of its peak value Opeak within less than two hours and the final value (after 10hours) (called Oend) must be lower than $0.1 \times \theta$.
\item \textbf{Oscillations:} After the initial peak the amplitude (difference between the local maximum and local minimum) must exceed $0.1 \times \theta$ at least 4 times.
\item[3.1]  \textbf{Dampened oscillations:}  The fifth amplitude must be less than half the second amplitude. 
\item[3.2]  \textbf{Sustained oscillations:} The fifth amplitude must be higher than half the second  amplitude. 
\end{enumerate}
We characterized the long-term behaviour of oscillations based on the relative amplitudes of the second and fifth peak because the first peak can be particularly high (Fig.~\ref{Fig2}D), and most dampened simulations have no more than five peaks. \\[0.3cm]

Quantitative data on the physiological concentrations of the cellular proteins and the transcription factor complex (nuclear Smad/CoSmad complex) do not exist, and we therefore had to set our detection threshold arbitrarily to  $\theta = 10$pM when analysing a unique constant stimulus with 200 pM TGF-$\beta$ ligand. When the response to several ligand concentrations or with several protein concentrations was studied we used the maximal response value as $\theta$. Simulations were run for 10 hours. In case if oscillations, if the amplitude of oscillations was still larger than $0.1 \times \theta$ after 10 hours, the simulation was continued until the oscillations vanished, but for a maximum 100 hours. In this way we avoid any impact of period length on the classification of oscillations, and the length of the period indeed does not bias our characterisation of oscillations to dampened or sustained oscillatory behaviour (Additional file 2, Fig.~S5C). The time thresholds 2 hours and 10 hours were chosen based on experimental data \cite{Nicolas2003}. \\[0.3cm]

\section*{Results and Discussion}
In response to a sustained stimulus (200 pM ligand) our simple model for TGF-$\beta$ signaling can give rise to sustained (Fig.~\ref{Fig2}A), transient (Fig.~\ref{Fig2}B), or oscillatory (Fig.~\ref{Fig2}C,D) responses. The sustained/transient distinction is particularly relevant, as it has been shown that those two qualitative behaviors are related to the growth inhibitory effect of TGF-$\beta$ \cite{Nicolas2003}. To better understand the conditions for these different behaviours we sought to identify parameter families that would give rise to a certain response type. We hoped that a comparison of those families would reveal the critical parameters that determine the response type. To that end we screened a large number of parameter sets and classified them according to their responses as described in detail in the Materials and Methods section. \\[0.3cm]

\subsection*{Parameter-dependent distinct qualitative responses}
Our criteria in Fig.~\ref{Fig2} are very strict (i.e. the speed of responses is an important criterium) and there is a wide undefined range between sustained and transient responses. As a consequence most parameter sets ($57.5\%$) do not fall into any of the defined categories (Additional file 2, Fig.~S5A, black sets). Of those that can be classified most ($25.5\%$ of the parameter sets tested) led to no response (Additional file 2, Fig.~S5A, marked in grey). Among the ''responsive" parameter sets most lead to sustained responses ($14.8\%$ of the parameter sets tested) (Additional file 2, Fig.~S5A, marked in red) while transient responses are observed less frequently ($2.2\%$of the parameter sets tested) (Additional file 2, Fig.~S5A, marked in green). All three behaviors have previously been observed in various models of TGF-$\beta$ signaling. We find that in addition in a minority of cases (306 simulations, i.e $0.046\%$) also oscillatory responses can be produced (Additional file 2, Fig.~S5B). Even though the number of sets that give rise to oscillations in the concentration of nuclear transcription factor complexes is small, these may occupy a sufficiently dense subspace in the parameter space to be physiologically relevant. The oscillations can either be sustained or dampened, depending on how fast their amplitude decays (Fig.~\ref{Fig2}C,D). As expected sustained oscillations have a larger number of peaks (Additional file 2, Fig.~S5B). While the period of the oscillations is not biased to dampened or sustained oscillatory behaviour (Additional file 2, Fig.~S5C), the duration (the time until oscillations vanish) depends on both the number of peaks and the duration that tends to be higher for sustained oscillations (Additional file 2, Fig.~S5D). \\[0.3cm]

Oscillatory behavior has been reported for a number of other signaling pathways (i.e. the ERK cascade \cite{Shankaran2009}), but so far no experimental evidence exists for oscillations in the TGF-$\beta$ pathway. However, standard biochemical experiments average over a large number of non-synchronized cells. If the nuclear concentration of transcription factor indeed oscillated, only sophisticated single-cell assays would reveal these. \\[0.3cm]

\subsection*{The impact of kinetic parameters on the response type}

We wondered which kinetic parameters would be critical for the different response types. Our sampling space is huge (23 parameters, with most of them sampled over 4 orders of magnitude, Additional file 1, Table S1)  and we looked for parameters that would be constrained in the different response types. In Fig.~\ref{supp1} we plot the sampled ranges in grey, and the parameter ranges that correspond to the different response types in colours. Since we are sampling from a uniform logarithmic distribution parameters that are not affecting the response type should remain uniformly logarithmically distributed in the parameter subsets. In Fig.~\ref{supp1}A we compare the parameter ranges of sustained (red) and transient (green) responses. We notice that whereas some parameter values remain (almost) uniformly distributed, others are constrained. Constrained parameters include the rates that describe the I-Smad dependent negative feedback loop (parameters $k5$, $k6$, $k14$, $k15$, $k17$, $k18$, and $k19$), the shuttling rate between cytoplasm and nucleus ($k8$), the dynamics of the Smad homo- and heterodimer formation/dissolution($k10$,and $k11$), and the dephosphorylation of Smad ($k13$). Fig.~\ref{Fig4}A shows the clear segregation of the ''sustained" (red) and ''transient" (transient) parameter sets in a plane spanned by the parameters that determine the strength of the negative feedback ($(k14\times k18) / (k15\times k17 \times k19)\times k5/k6$) and the speed of Smad dephosphorylation, $k8\times (k11\times k13)/k10$. To favour transient responses over the sustained responses the I-Smad dependent negative feedback must be strong and dephosphorylation of Smad must be fast. The need for rapid dephosphorylation likely arises also because of our requirement that adaptation must happen within 2 hours. We notice that the size of the parameter set that permits transient responses is considerably smaller than the parameter set that permits sustained responses. However, transient responses can also result from degradation of core signaling components and ligand which is not considered here. \\[0.3cm]

Transient and oscillatory responses are similar in that the response must decay quickly in spite of the continuous presence of ligand. A  similar comparison of the parameter ranges that permit transient (green) or oscillatory (blue and magenta) responses (Fig.~\ref{supp1}B) indeed reveals that similar restrictions apply (i.e. large shuttling rate between cytoplasm and nucleus, $k8$, and strong negative feedback, $k14, k15,  k17, k18$, and $k19$). However, in case of oscillations the response restarts and in addition we indeed notice a strong restriction of the rate of ligand-receptor binding $k2$ in case of oscillatory responses. Fig.~\ref{Fig4}B shows the clear segregation of the parameter sets that give rise to ''transient" (green), dampened (blue) or sustained (purple) oscillatory responses in a plane spanned by the receptor-ligand binding rate $k2$ and the speed of I-Smad turn-over ($k16\times k17\times k19$). Oscillations are observed only when $k2$ is small such that ligand binds slowly to its receptor and the pool of free receptor is depleted gradually (Additional file 2, Fig.~S6A). As a consequence free receptor is still available when I-Smad has downregulated the response and ligand can still trigger a further response. Conversely, if $k2$ is large, receptors are rapidly bound to the ligand (whose concentration is constant in this model), and once the response has terminated, there is no free receptor available to induce a new response (Additional file 2, Fig.~S6B). $k2$, and thus the speed with which the free receptor concentration decreases, critically determine the dampening of oscillations. A 10-fold change in the value of $k2$ can transform sustained oscillations in highly dampened ones (compare panels A and B in Additional file 2,  Fig.~S6). Rapid degradation of I-Smads is important for sustained oscillations because otherwise all receptors become rapidly sequestered by I-Smad and the response is terminated. Accordingly inclusion of receptor endocytosis and recycling to the membrane (cycling) combined with the removal of the I-Smad would allow further oscillatory cycles.  \\[0.3cm]

Each parameter in our simple model integrates the effects of many further interactions as may also arise from cross-talk. Thus it has been shown that the phosphorylation of Smad in its linker region by Ras-activated MAPK induces a cytoplasmic retention of R-Smads \cite{Kretzschmar1999}, which in our system would be represented by a lower shuttling rate into the nucleus ($k8$). Interestingly $k8$ indeed strongly influenced the response type. Another parameter that appears to be important in determining the response characteristics is the binding rate of TGF-$\beta$ to the receptor ($k2$). Our description of the processes at the cell membrane is very simple and thus $k2$ has also to take in account the regulation of TGF-$\beta$ outside the cell by soluble sequestering factors and membrane-bound co-receptors as well as processes that affect the receptor density on the cell membrane. Those auxiliary factors play therefore a crucial role in the TGF-$\beta$ pathway flexibility. We should stress that all parts of the parameter space should be readily reachable for the cell and small adjustments in the parameter values should thus be sufficient to alter the response type. \\[0.3cm]

\subsection*{The regulatory impact of cellular protein concentrations}
 
The kinetic rate constants of a reaction depend on the particular protein chemistry. While rate constants may be different between species, rate constants are unlikely to differ between individuals of one species and even more unlikely to differ within a single individual. However, during the development of an organism the same signaling network can elicit qualitatively  different responses at different times and locations. We therefore wondered whether changes in the protein concentrations (which can be easily adjusted by an organism or result from crosstalks with other signaling pathways) would enable the required regulatory flexibility. To find parameter ranges that would permit such flexibility we repeated our previous screen with different concentrations of receptors, R-Smad or Co-Smad: for each of the three species we first carried out 3 screens where concentrations were increased or decreased from their reference concentration $c_0$ to $c_0/100$ or $c_0\times 100$. We then looked for parameter sets that would permit a switch between a transient and a sustained output response as the protein concentrations changed. Our parameter sampling space is huge (23 parameters, with most of them sampled over 4 orders of magnitude, Additional file 1,  Table S1) and a switch could be observed for less than $1\%$ of the sets (Fig.~\ref{Fig5}A, black bars). When we plotted the parameter ranges for which we observed switching we noted that a number of parameter ranges were restricted compared to the initial sampling range (Additional file 2, Fig.~S7). We therefore wondered whether there would be particular parameter ranges for which concentration-dependent switching would be more frequent. Indeed when we reduced the sampling ranges of the parameter values (Additional file 2, Fig.~S7) about $20\%$ of the parameter sets enabled switching as the R-Smad concentration was varied, $25\%$ as the receptor concentration was varied, and almost  $30\%$ as the Co-Smad concentration was varied (Fig.~\ref{Fig5}A, grey bars). We notice that the only rates that were not restricted while enhancing the fraction of parameter sets that permit switching were the rate of ligand-$TGF\beta R$ unbinding ($k1$) and of I-Smad mRNA export ($k16$). These rates thus appear to have very little influence on the overall kinetics within the screened range. We next wondered what would be the minimal change needed in protein concentration to allow the switch. To that end we carried out 9 supplementary screens for each of the three species where concentrations were increased or decreased from their reference concentration $c_0$ over a 100-fold range in multiples of 3, i.e. $c_0(n) = c_0 3^n$ with $n = [-4, -3, \ldots, 3,4]$. Interestingly, while a change in the response type was observed most frequently in response to changes in the Co-Smad concentration (Fig.~\ref{Fig5}A, grey bars), switches could be achieved  with much smaller concentration changes when the receptor or R-Smad concentration were varied (Fig.~\ref{Fig5}B). Thus only a 3-10 fold change in  the receptor and R-Smad concentration was typically required while the Co-Smad concentration typically needed to be changed by 20-100-fold. The I-Smad Smad6 has indeed been reported to inhibit TGF-$\beta$ signaling by sequestering the Co-Smad Smad4 in an inactive complex \cite{Hata:1998vx}.  It has further been argued that cross-talk between different TGF-$\beta$ pathways may be integrated via a competition for Co-Smads. Based on our observations such competition would need to greatly alter the concentration of available Co-Smad to be effective and the receptor and the R-Smad would  offer a more sensitive point of control. Previous models have focused on the dynamical control of the TGF-$\beta$ receptor and have shown that this indeed offers great regulatory flexibility \cite{Vilar2006}. Experiments further show that the I-Smad may also affect the turn-over rate of R-Smads and thus affect their cellular concentration \cite{Wu:2009p8501}. \\[0.3cm]

\subsection*{TGF-$\beta$ dose-dependent response}

Finally we wondered how different ligand concentrations would affect the cellular response. The impact of different TGF-$\beta$ concentrations have already been studied by Clarke \textit{et al.} \cite{Clarke2009} and Zi\textit{ et al.} \cite{Zi2011}, but there the results were strongly affected by ligand depletion since the TGF-$\beta$ concentrations were allowed to go down over time because of internalization and degradation. We were interested how different, but constant, stimuli would affect the response - the effect of ligand depletion can then be deduced as response to decreasing ligand concentrations. As we varied the ligand concentration between 0.2 pM and 20 nM we noticed that only for a very small fraction of parameter sets (less than $0.5\%$) the response type changed qualitatively as the concentration varied. The parameter sets were not clustered and a further increase by restricting parameter ranges (as in case of the cellular protein concentrations) could not be achieved. Even though changes in the TGF-$\beta$ concentration cannot switch the response type in our simulations, the duration of the response increases with increasing TGF-$\beta$ concentrations as previously observed by Zi \textit{et al.} \cite{Zi2011}. This increase was, however, insufficient to alter the response type according to our definitions.  \\[0.3cm]

The ligand concentration clearly affects the maximal response in our simulations, and the transcription factor activity increases with the ligand concentration until a plateau is reached (Fig.~\ref{Fig6}A). In case of sustained responses (but not for transient responses) the peak value is reached more quickly at higher ligand concentrations (data not shown). The saturation curve in Fig.~\ref{Fig6}A can be fitted with an exponential curve, i.e. $O_{peak} = \max{(O_{peak})}(1- \exp{(-x / \eta)})$ where $x$ refers to the TGF-$\beta$ ligand concentration ($x =0.2, 2, 6, 10, 15, 20, 50, 100, 150, 200, 1000, 20000$ pM) and the parameter $\eta$ indicates the concentration range for which the response saturates. Histograms of $\eta$ (Fig.~\ref{Fig6}B,C) show that the sustained response (Fig.~\ref{Fig6}C) tends to saturate at lower TGF-$\beta$ concentrations than transient responses (Fig.~\ref{Fig6}B). Moreover, in case of sustained responses there is a biphasic distribution in the saturation concentrations with one peak around 0.1 pM and the other one around 10pM (Fig.~\ref{Fig6}C). However, in both transient and sustained cases, the transcription factor is able to reach similar maximal values (Additional file 2, Fig.~S8). On the contrary, the maximal output value reached by oscillating responses is much lower than in the sustained and transient case. Our results are mostly in agreement with the conclusions drawn by Chung \textit{et al.} (2009), who showed also that transient TGF-$\beta$ responses saturate. However, deviating from our results, Chung and co-workers observed that also in transient responses the peak value is reached more rapidly as the stimulus concentration increases.  \\[0.3cm]

For  parameter sets that give rise to oscillatory responses, changing the input strength and shape does not influence the period of oscillation but modulates the evolution of the oscillations amplitudes (data not shown). When exposed  to sustained, high TGF-$\beta$ concentrations the amplitude of oscillations starts to decay from the beginning. When the TGF-$\beta$ concentration raises progressively, the amplitude of oscillation first raises and then decays, reflecting two competing phenomena : the amplitude of oscillations tends to be proportional to the input, but at the same time the sequestration of the receptor by the inhibitor leads to a dampening of the amplitude. \\[0.3cm]

We next investigated in how far the kinetic parameters can influence the saturation concentration (Fig.~\ref{Fig6}B,C) and the maximal output value at saturation (Additional file 2, Fig.~S8 and Additional file 2, Fig.~S9). For transient responses it is mainly the rate of ligand-receptor binding, $k2$, that determines the saturation concentration (Fig.~\ref{Fig6}D and Additional file 2, Fig.~S10). In case of slow binding higher concentrations of ligand are required to saturate the receptors. The saturation concentration for sustained responses are determined both by the receptor-ligand binding rate, $k2$, and by the cytoplasm-nucleus shuttling rate, $k8$ (Fig.~\ref{Fig6}E and Additional file 2, Fig.~S10). Fast shuttling enables more rapid deactivation of Smads as based on observations by Hill and coworkers \cite{Schmierer2008} dephosphorylation is restricted to the nucleus in our model. As discussed above $k2$ and $k8$ have both been reported to be modulated by other processes. The saturation concentration can therefore also be adjusted by cross-talk.  \\[0.3cm]

The different saturation concentrations are likely important for the TGF-$\beta$ response as different genes can be activated or repressed depending on the nuclear Smad complex concentration. While the mechanism by which different concentrations of the nuclear transcription factor complex translate into different transcriptional responses has not been resolved, likely mechanisms include promotor selection based on differences in the promoter binding-site affinities, cross-repression, and the establishment of a reciprocal of repressor gene expression \cite{Schmierer2007, Ashe2006}.

\subsection*{Proportional  "faithful" responses}

When ligands of the TGF-$\beta$ family act as a morphogen, as it is for example the case for Dpp in\textit{ Drosophila} or Activin in \textit{Xenopus}, cells must finely sense extracellular concentrations and transduce this signal inside the cell. We therefore looked for parameter sets leading to a response proportional to the input which we term "faithful". The parameter sets that gave rise to anything but sustained responses (i.e. transient, oscillatory, non-responsive, undefined responses) to sustained ligand exposure can already be discarded. Those parameter sets that gave rise to sustained responses to sustained ligand exposure we sought to analyse further with dynamic input signals.  Here we used as input a function that first linearly increased from 0 to 720 pM for 5 hours and then linearly decreased to zero over the next 5 hours  (Fig.~\ref{Fig7}A). To screen our simulations for "faithful" parameter sets we normalized both the input and the output with respect to their respective highest values, and calculated the squared residuals $R$ between input and output according to $R=\displaystyle\sum_{j}^{}(\textrm{input}_{j}-\textrm{output}_{j})^2$. The $10\%$ sets with the lowest residual were classified as "faithful" and the  $10\%$ sets with the highest residual were classified as "unfaithful" for further analysis (Additional file 2, Fig.~S11). \\[0.3cm]

A response is faithful if the output is proportional to the input over time, i.e. $y_{output}(t)=\alpha \times y_{input}(t)$, where $\alpha$ is the proportionality coefficient. This requires (i) that the output adapts rapidly to changes in the input, and (ii) that the response does not saturate, i.e. $\max{(y_{output}(t))} < max(O_{peak})$, which is the case if the proportionality coefficient $\alpha$ is low and/or the maximal response value $max(O_{peak})$ is high. Those requirements are reflected in the constraints on the parameter values (Fig.~\ref{Fig7}B-C) for faithful responses, i.e. a low binding rate of TGF-$\beta$ to its receptor and a low phosphorylation rate prevent early saturation of the output, while a relative weak feedback and a low binding rate of the I-Smad to the receptor prevent a premature termination of the response. We have previously discussed the regulation of the binding rate of TGF-$\beta$ to its receptor, $k2$ and thus now focus on the feedback. The different I-Smads have been shown to vary in their effects. Thus Dad, the \textit{Drosophila} I-Smad, appears to interfere mainly with the BMP-like pathways (Tkv and Sax receptor dependent pathways) but not the Activin-like Babo-dependent pathway \cite{Kamiya:2008p565}.  Inhibition by vertebrate Smad6 and Smad7 can be achieved by sequestration, enhanced degradation, or an impact on phosphorylation. The different processes likely have different efficiencies and this will determine the efficiency of the negative feedback. \\[0.3cm]

Our results indicate that under certain parameter restrictions the extracellular concentration is directly reflected in the output concentration. In that case, TGF-$\beta$ can act as a morphogen, conveying positional information and determining cell-fate, subjected to the set of activated and repressed genes.

 \section*{Conclusions}

The duration of the signaling response is thought to be an important factor influencing the cell's phenotypic response to 
TGF-$\beta$. We have employed a very simple model of the TGF-$\beta$ network to better understand the mechanistic basis of the observed signaling plasticity. We find that the qualitative response (transient, sustained, oscillations, proportional responses)  to a constant ligand exposure can indeed be changed by altering the value of a single parameter value. Since we consider a simple model each parameter value represents a wider range of processes and our observation thus implies that both changes in protein concentration as well as cross-talk between signaling pathways can alter the qualitative response to a TGF-$\beta$ stimulus. Many more complicated models for TGF-$\beta$ signaling as well as for other signaling networks have been proposed already. To better understand the regulatory impact of cross-talk it will be important to connect experimentally validated models for the TGF-$\beta$ network also to those for other pathway models. While many kinetic parameters have been measured an important parameter that remains often unmeasured is the protein concentrations. To better predict the responses in different cell types it will be important to obtain quantitative information on protein abundance in different cell types - and eventually in individual cells. \\[0.3cm]

\section*{Authors contributions}
DI and GF developed the model, GF and MH carried out an initial analysis of the model, GF developed the infrastructure for the parameter screens, GC carried out the analysis of the model, MH created Fig.~1, GC and DI wrote the paper. All authors read and approved the final manuscript.

\section*{Acknowledgements}
  \ifthenelse{\boolean{publ}}{\small}{}
This research was supported by an iPhD SystemsX grant to Georgios Fengos.


\newpage
{\ifthenelse{\boolean{publ}}{\footnotesize}{\small}
 \bibliographystyle{bmc_article}  
\bibliography{./biblio}


\ifthenelse{\boolean{publ}}{\end{multicols}}{}



\newpage 
\section*{Figures}

\begin{figure}[!ht]
\begin{center}
\includegraphics[width=15cm]{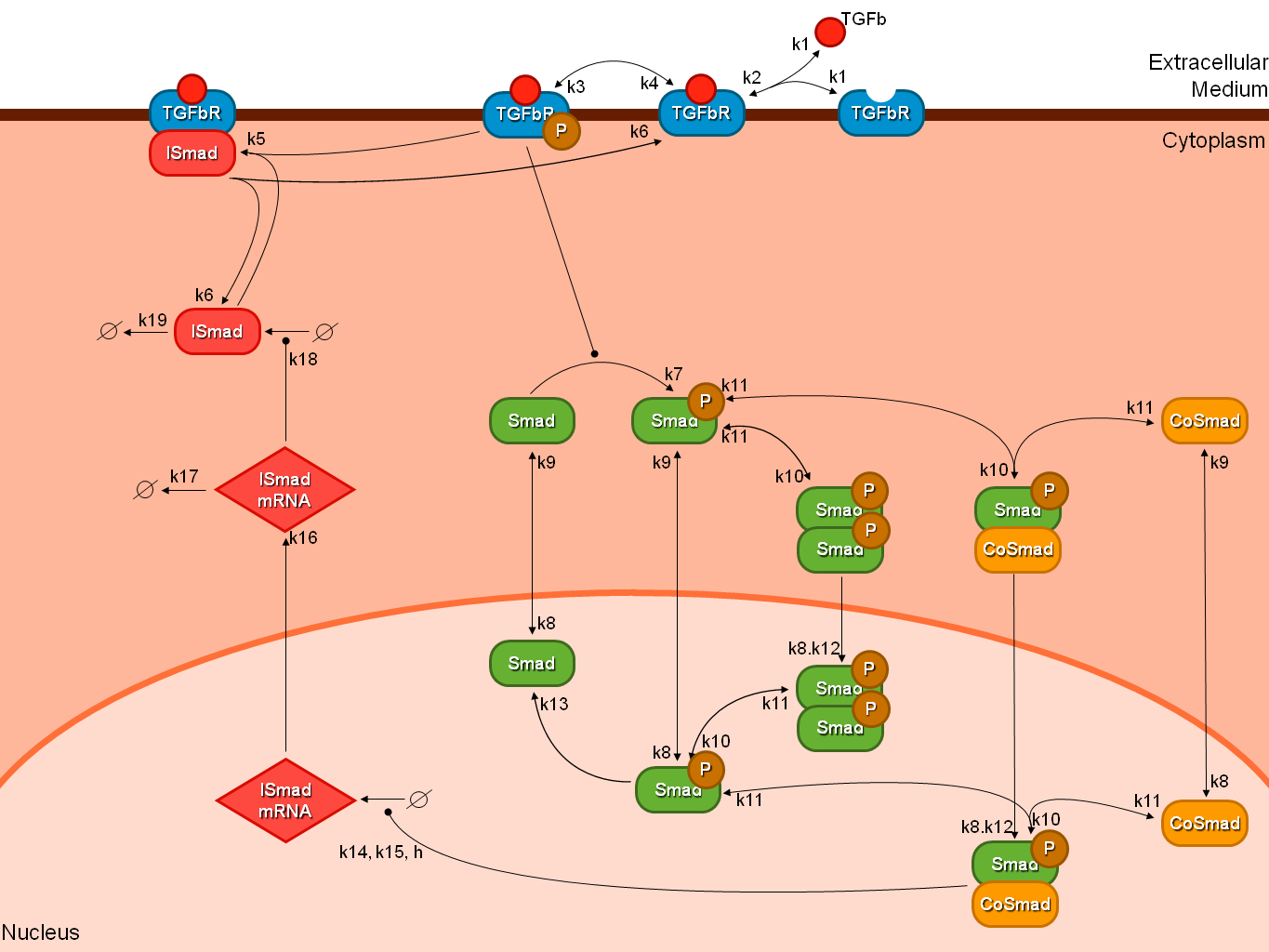}
\end{center}
\caption{
{\bf  A simple model of TGF-$\beta$ signaling with I-Smad mediated negative feedback. }The ligand TGF-$\beta$ reversibly binds to the TGF-$\beta$ receptor (reactions 1 and 2), which is then phosphorylated to become fully active (3 and 4). The active receptor induces phosphorylation of R-Smad (denoted simply Smad)(7), which in turn can reversibly dimerize or form a complex with Co-Smad (10 and 11). Those two reactions can take place either in the cytoplasm or in the nucleus and the five species Smad, phosphorylated Smad, Co-Smad, homodimers and heterodimers can shuttle from the cytoplasm to the nucleus and back (8, 9 and 12). Nuclear Smad/Co-Smadf complexes act as transcription factors and trigger the transcription of I-Smad mRNA in the nucleus (14 and 15). The I-Smad mRNA then shuttles to the cytoplasm (16), where it can be degraded (17) or translated into I-Smad (18). I-Smad mediates a negative feedback by sequestering the active receptor (5 and 6) and can be degraded (19). 
}
\label{Fig1}
\end{figure}

\begin{figure}[!ht]
\begin{center}
\includegraphics[width=15cm]{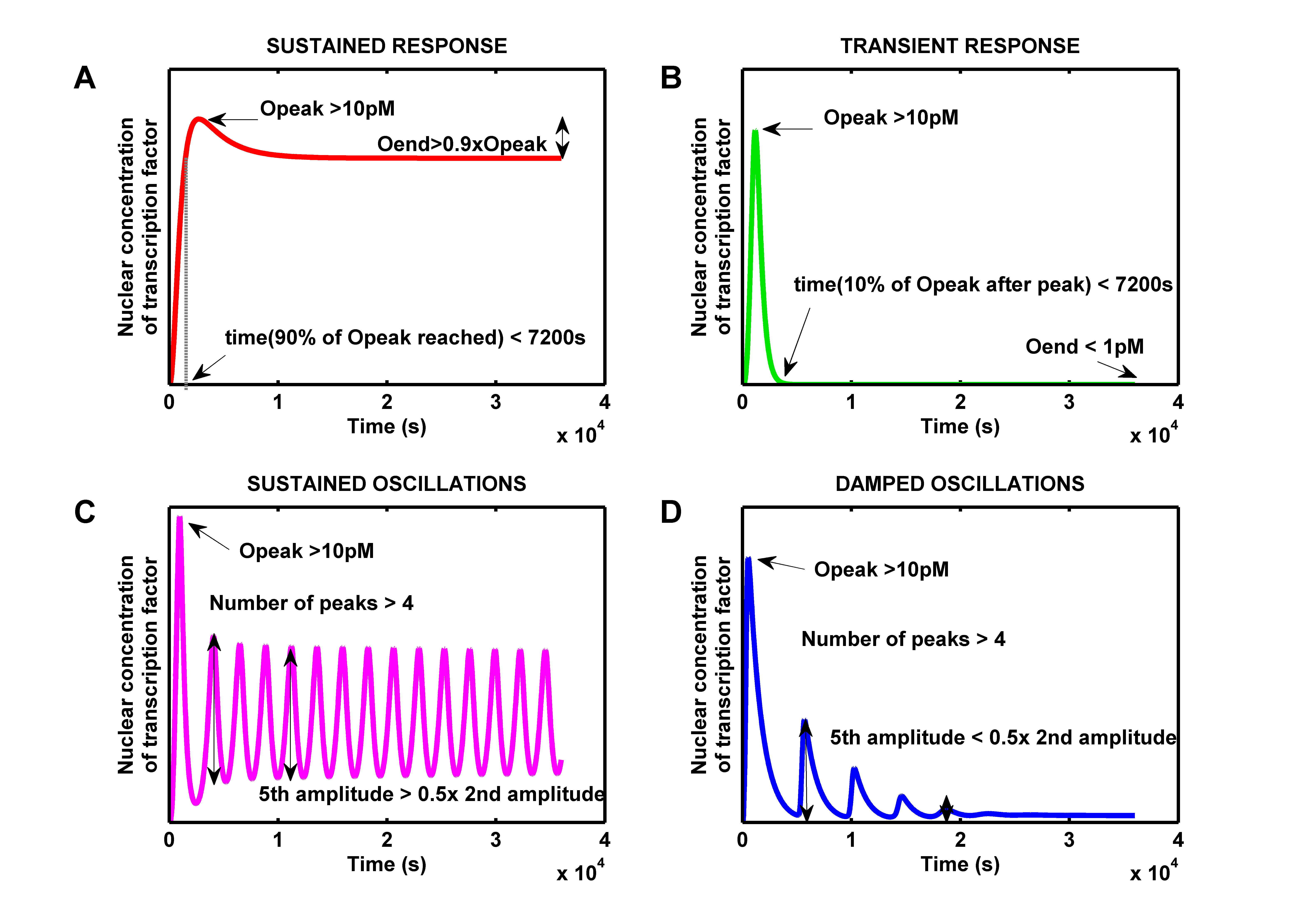}
\end{center}
\caption{
{\bf Criteria to define the different TGF-$\beta$ signaling responses.} All responses must exceed a threshold concentration of  $10$ pM initially to be considered responsive. \textbf{(A)} Sustained responses: The response must reach $90\%$ of the maximal value Opeak within 2 hours and retain $90\%$ of this maximal value until the end of 10 hours simulation. 
\textbf{(B)} Transient responses: The response must exceed 10 pM and subsequently return to less than $10\%$ of the highest value Opeak within 2hours of stimulation. The final value Oend must be lower than 1pM.
 \textbf{(C-D)} Oscillations: After the initial $\geq$10pM peak at least four further peaks must exceed 1 pM in amplitude. Depending on whether the fifth amplitude is less or higher than half the second amplitude we distinguish \textbf{(C)} sustained and  and \textbf{(D)} dampened  oscillations respectively. 
}
\label{Fig2}
\end{figure}

    \begin{figure}[!ht]
\begin{center}
\includegraphics[width=15cm]{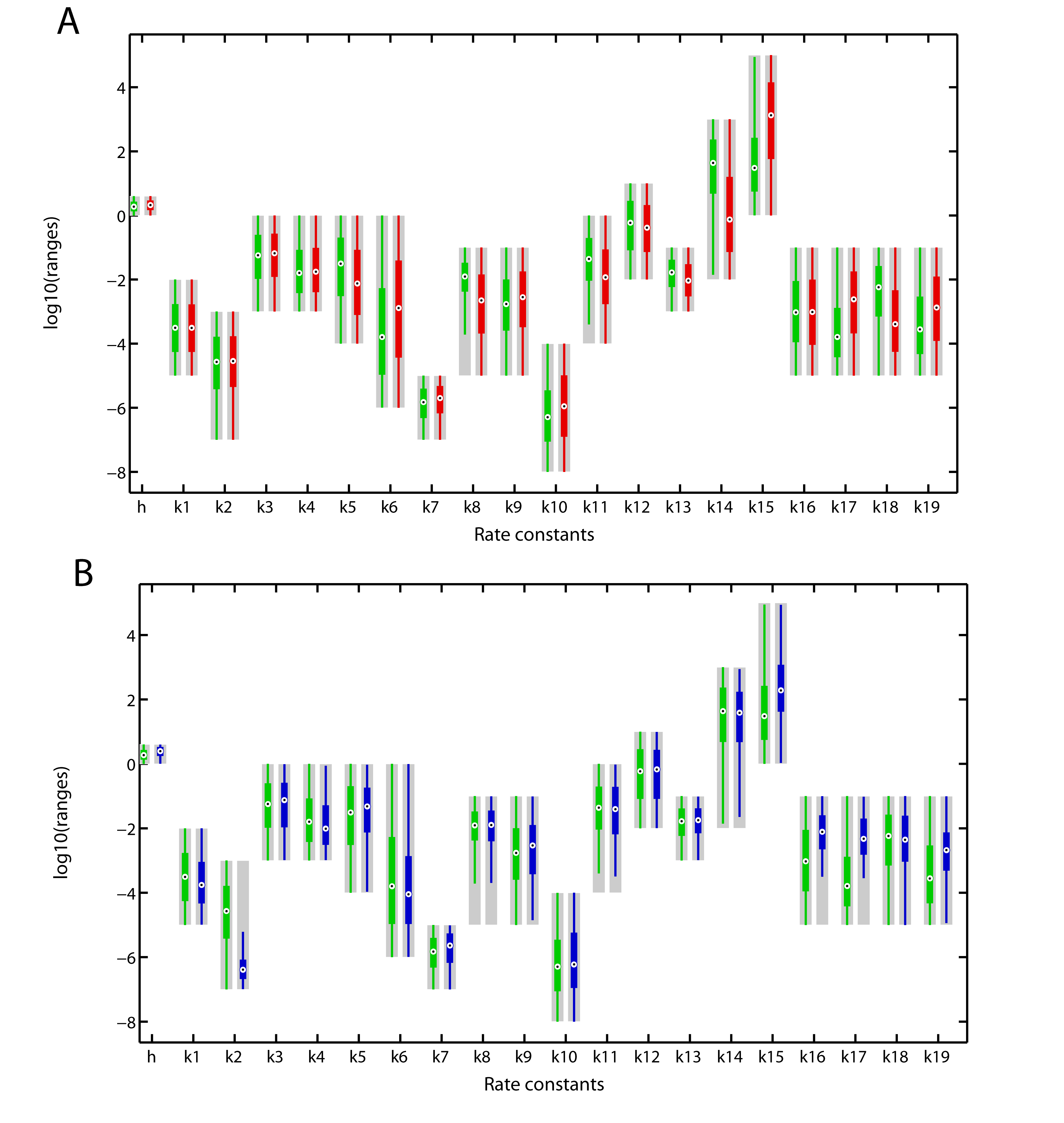}
\end{center}
\caption{
{\bf Box plots of the parameters corresponding to the different responses types} \textbf{(A)} Box plots of parameter sets leading to a transient (green) or sustained (red) response. Parameters that differ are mainly $k8$, $k10$, $k11$, $k13$ (shuttling rate from cytoplasm to nucleus, formation/dissolution of the Smad dimers, and dephosphorylation of R-Smad), and $k5$, $k6$, $k14$, $k15$, $k17$, $k18$, $k19$ (all related to the strengh of the feedback). \textbf{(B)} Box plots of parameter sets leading to a transient (green) or oscillatory (blue) response. $k16$, $k17$, $k19$ (dynamics of the I-Smad mRNA and I-Smad protein) and $k2$ (binding of TGF-$\beta$ to its receptor) are key determinants of the response kind. Ranges of the uniform sampling distributions, as stated in Table S1, are indicated by grey boxes. 
}
\label{supp1}
\end{figure}

\begin{figure}[!ht]
\begin{center}
\includegraphics[width=15cm]{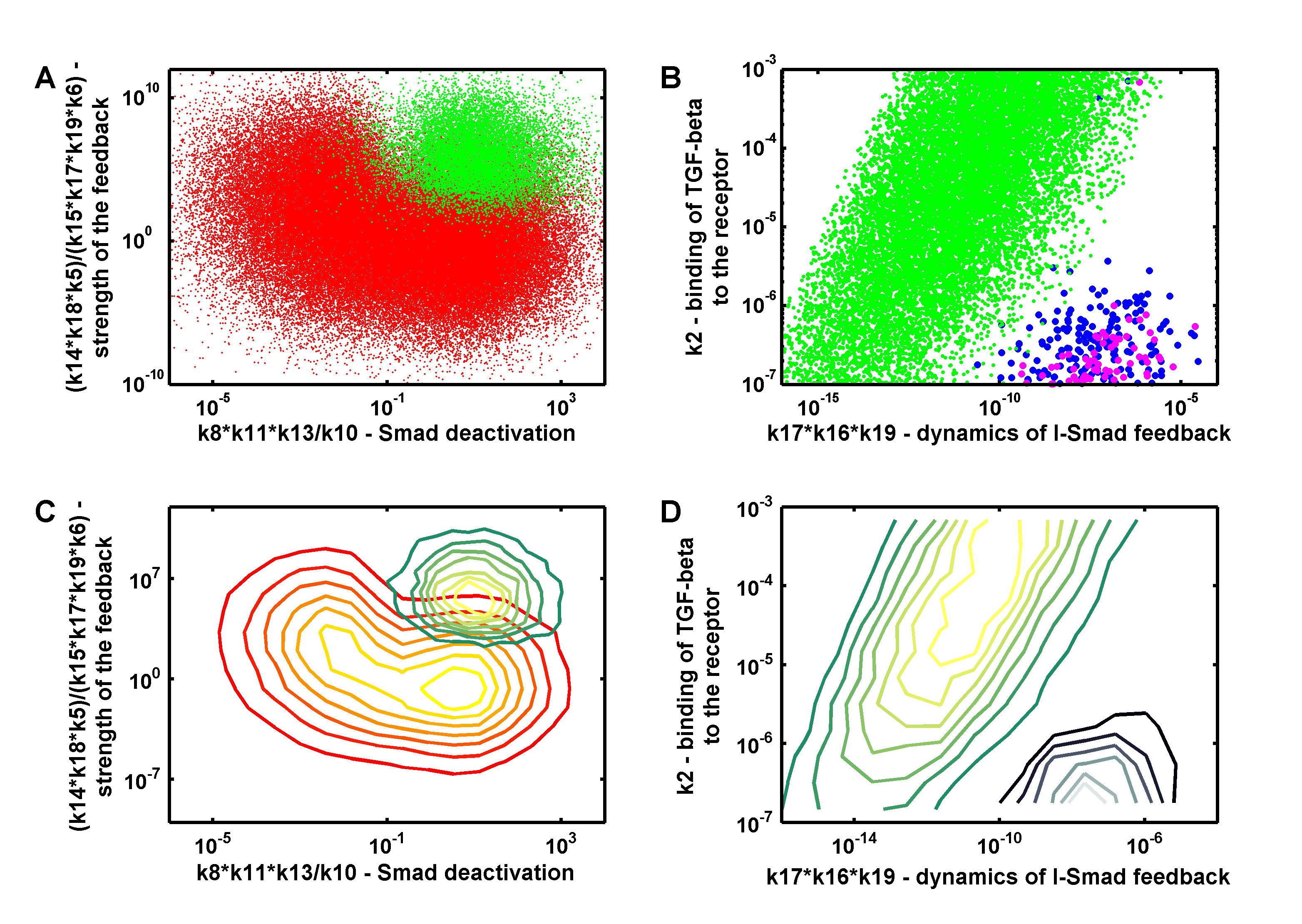}
\end{center}
\caption{
{\bf Impact of kinetic parameters on the type of the TGF-$\beta$ response.} \textbf{(A,C)} A strong negative feedback,  fast nuclear shuttling of Smads and a rapid dissociation of the dimers favour a transient (green) over a sustained response (red). \textbf{(B,D)} Fast production and degradation of the I-Smad mRNA and I-Smad protein is required  for oscillations to appear, and a low TGF-$\beta$ receptor on-rate enhances oscillatory (blue and magenta in the scatter plot, and black to blue in the contour plot) relative to a transient (green in the scatter plot, and green to yellow in the contour plot) response.
}
\label{Fig4}
\end{figure}

\begin{figure}[!ht]
\begin{center}
\includegraphics[width=15cm]{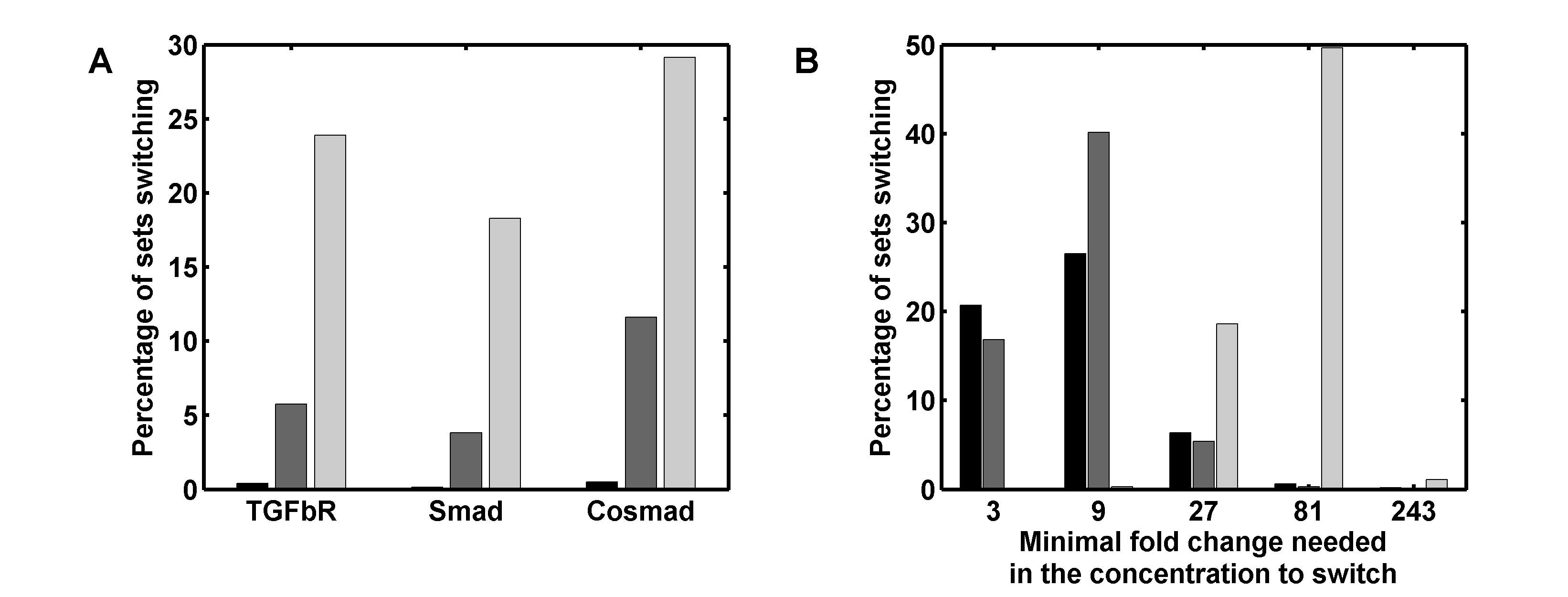}
\end{center}
\caption{
{\bf A change in protein concentrations can switch the type of the TGF-$\beta$ response.} 
\textbf {(A)} Percentage of parameter sets that permit a switch in the qualitative response (transient versus stustained response) to ligand when TGF-$\beta$ Receptor, R-Smad or Co-Smad concentrations are increased or decreased by 100-fold. The parameter sets were drawn from the ranges as specified in Additional file 1, Table S1 (black), or in more (grey) and more (light grey) restricted ranges as shown in Additional file 2, Fig.~S7. \textbf{(B)} The minimal relative change that is required in the concentrations of TGF-$\beta$ Receptor (black), R-Smad (grey), or Co-Smad (light grey) to switch between transient and sustained responses when parameters were drawn from the most restricted range (corresponding to the light grey column in panel A and the lightest colour in Additional file 2, Fig.~S7.
}
\label{Fig5}
\end{figure}

\begin{figure}[!ht]
\begin{center}
\includegraphics[width=15cm]{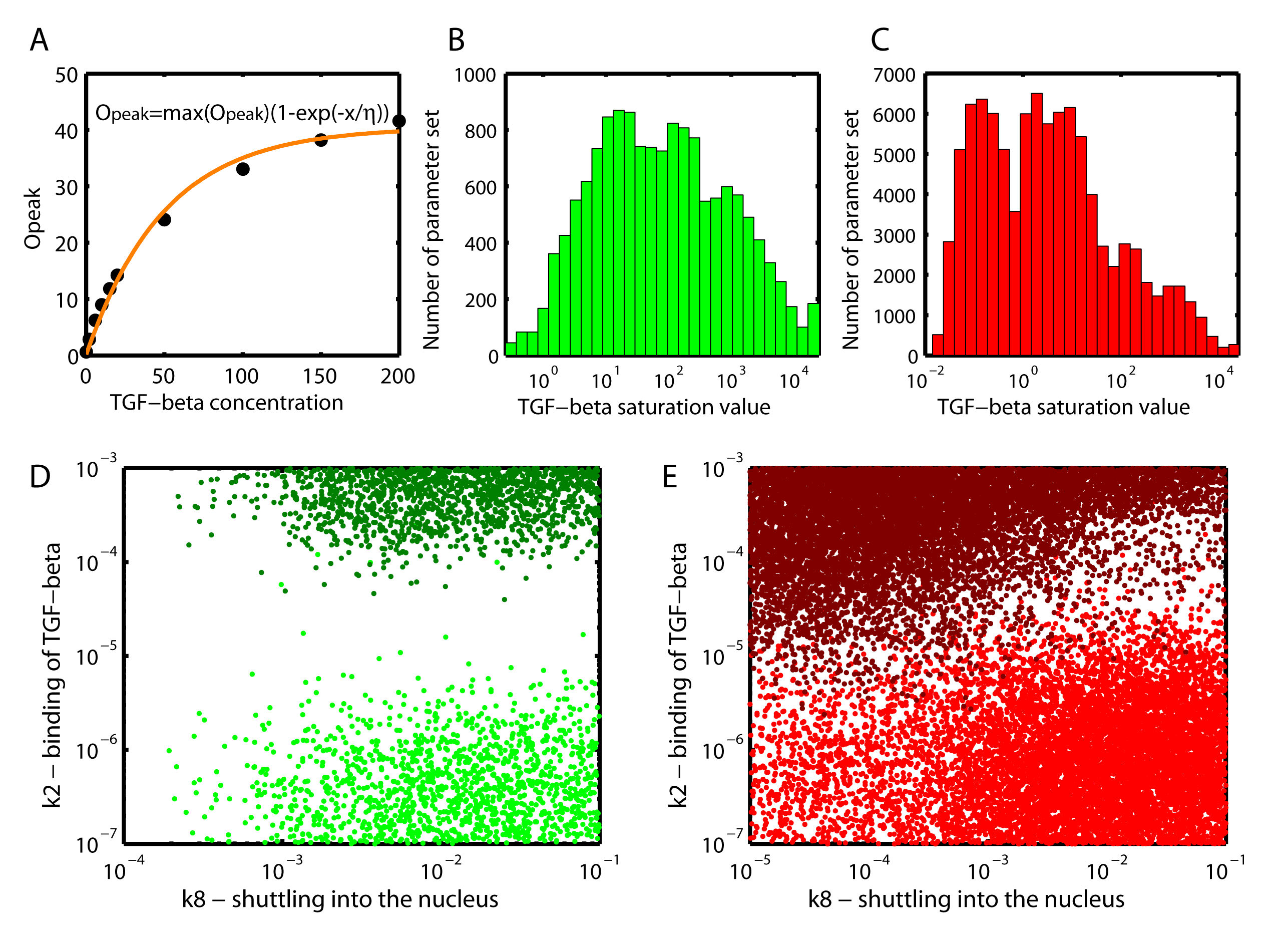}
\end{center}
\caption{
{\bf TGF-$\beta$ dose-dependent response.} 
\textbf{(A)} The pathway response increases with increasing ligand concentration until a plateau is reached. The saturation curve can be described with an exponential function, $O_{peak} = \max{(O_{peak})}(1- \exp{(-x / \eta)})$ where $x$ refers to the ligand concentration (which was sampled at 12 concentrations between 0.2 pM and 20 nM), and the parameter $\eta$ is characteristic for the saturation concentration. 
\textbf{(B,C)} A histogram of the distribution of $\eta$, the parameter characteristic for the saturation concentration for (B) the transient response, and (C) the sustained response. \textbf{(D)} For the transient set, the parameter that determines if the saturation concentration is high (green) or low (dark green) is $k2$, the binding rate of TGF-$\beta$ to its receptor. \textbf{(E)} For the sustained set, two parameters are crucial. $k2$ (binding of TGF-$\beta$) and $k8$ (shuttling into the nucleus) set the saturation concentration either to high (red) or low (dark red).
}
\label{Fig6}
\end{figure}

\begin{figure}[!ht]
\begin{center}
\includegraphics[width=15cm]{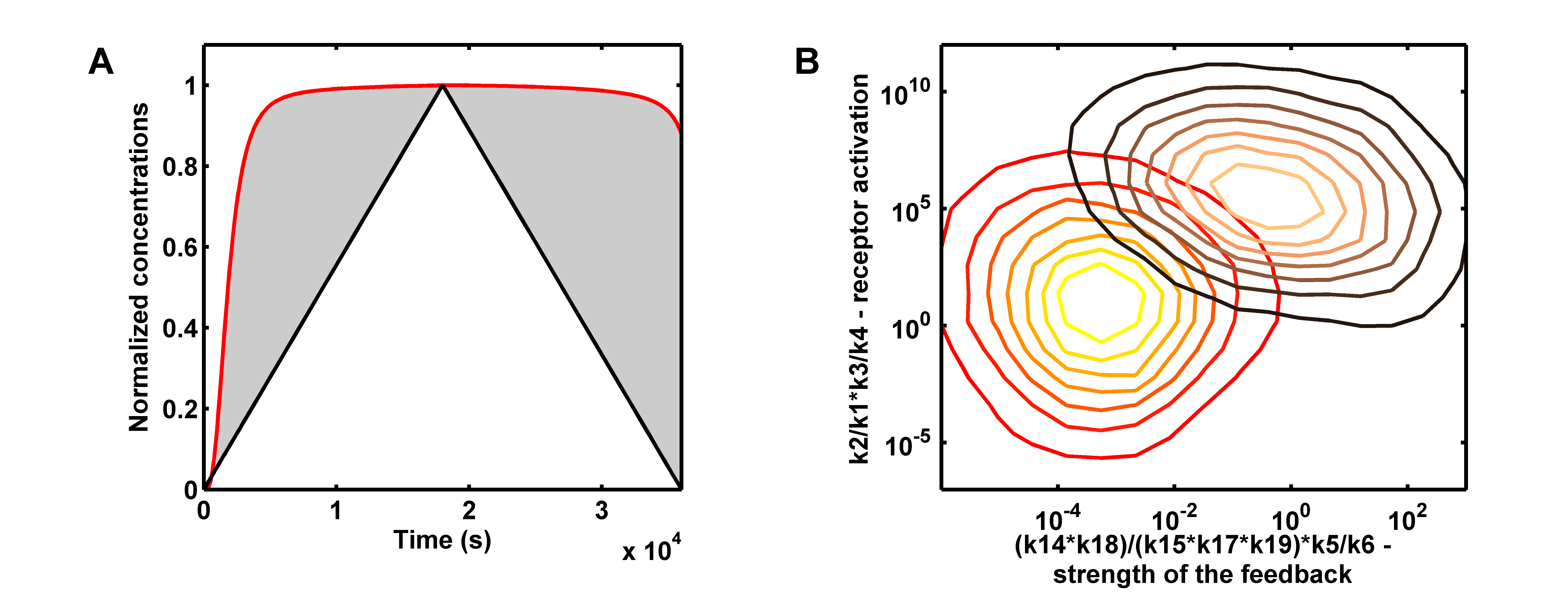}
\end{center}
\caption{
{\bf Parameter Dependency of Faithful Responses.} \textbf{(A)} To investigate the faithfulness of the response (red curve), we re-analysed those parameter sets that had produced sustained responses to a sustained input (red parameter sets in Additional file 2, Fig.~S5) with time-varying inputs (linearly increasing and then decreasing TGF-$\beta$ input concentration, black curve). Based on the  squared residuals (grey area) between the normalized inputs (black line) and outputs (red line) we defined faithful and unfaithful responses as those in the first and last 10-quantile respectively. 
\textbf{(B)} An inefficient activation of the TGF-$\beta$ receptor and a weak negative feedback favours faithful (light red to yellow) over unfaithful (black to salmon-pink) responses.
}
\label{Fig7}
\end{figure}



\newpage
\clearpage

    
    \renewcommand{\thefigure}{S-\arabic{figure}}
\setcounter{figure}{0}

\renewcommand{\theequation}{S-\arabic{equation}}
\setcounter{equation}{0}

\renewcommand{\thetable}{S\arabic{table}}
\setcounter{table}{0}

\subsection*{Additional files}

Additional file 1 \\
Title : Supplementary\_Tables \\
Description : Tables of the model parameters and the equations used in the model. \\[0.3cm]

Additional file 2 \\
Title : Supplementary\_Figures \\
Description : Figures for the response classification of the parameter screen (Fig.~S5), the dependance of the damping of oscillations on $k2$ (Fig.~S6), boxplots for the parameter sets that can lead to both transient and sustained responses (Fig.~S7), distribution of the maximal output value (Fig.~S8), boxplots for parameters with high and low saturation value (Fig.~S10), and with high and low maximal value (Fig.~S9), boxplots for faithful and unfaithful parameters sets (Fig.~S11), the evolution of the concentrations of all species in a representative cases (Fig.~S1, S2, S3 and S4).\\[0.3cm]

Additional file 3 \\
Title : Model\_in\_Cell\_Designer \\
Description : Systems Biology Markup Language (SBML) file of the model.

}

\end{bmcformat}
\end{document}


\renewcommand{\thefigure}{S\arabic{figure}}
\setcounter{figure}{0}


\title{Supplementary figures \\ The plasticity of TGF-$\beta$ signaling}
 

\author{Geraldine Celli\`ere, Georgios Fengos, Marianne Herv\'e and Dagmar Iber}
\date{}
\maketitle

\begin{figure}[!ht]
\begin{center}
\includegraphics[width=13cm]{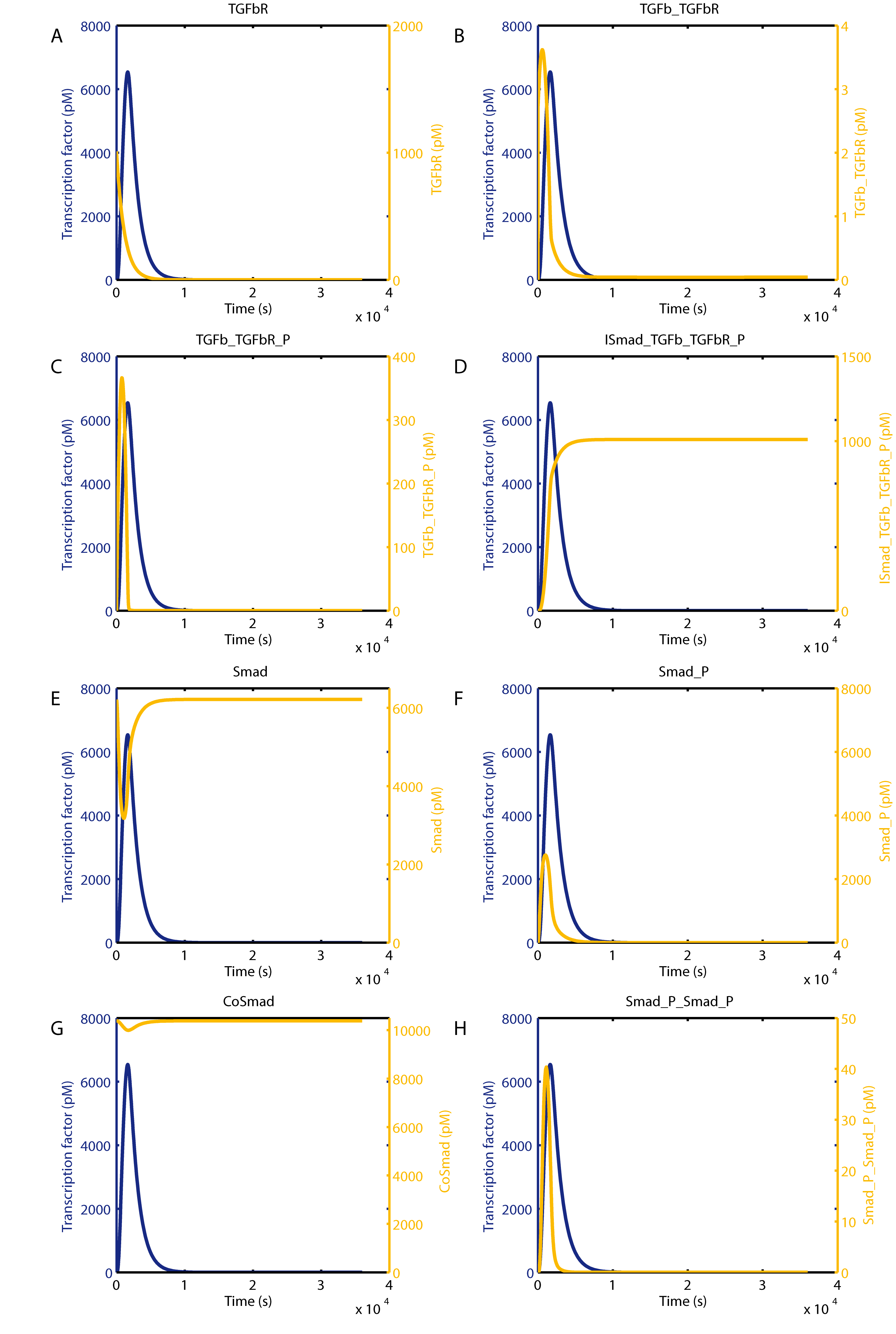}
\end{center}
\caption{
{\bf Evolution of the concentrations of all species in a representative transient response case.} Note that different scales are chosen for each graph. The parameter set used is listed in Additional file 1, Table S3.
}
\label{supp9}
\end{figure}

\begin{figure}[!ht]
\begin{center}
\includegraphics[width=13cm]{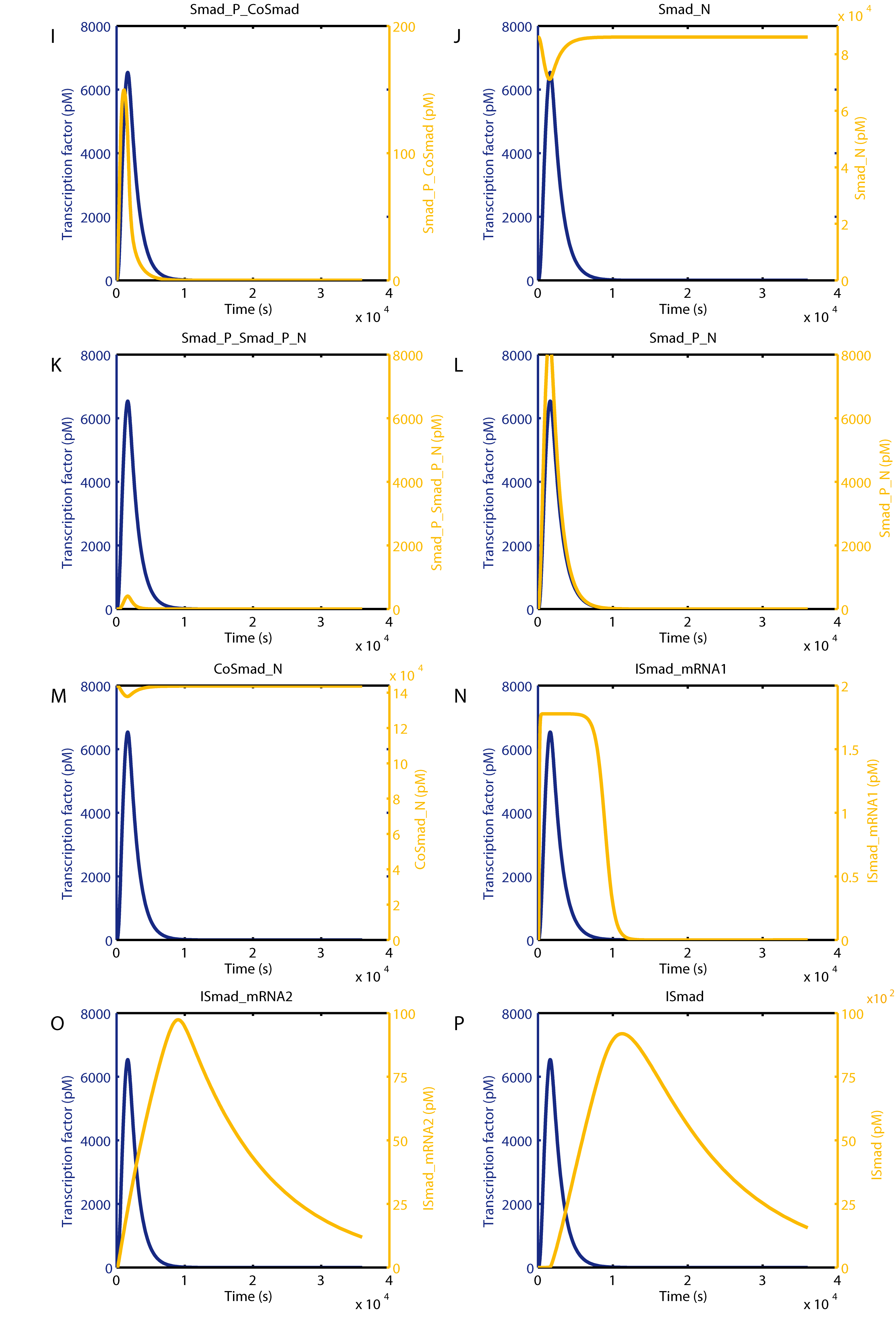}
\end{center}
\caption{
{\bf Evolution of the concentrations of all species in a representative transient response case (continued).} Note that different scales are chosen for each graph. The parameter set used is listed in Additional file 1, Table S3.
}
\label{supp10}
\end{figure}

\begin{figure}[!ht]
\begin{center}
\includegraphics[width=13cm]{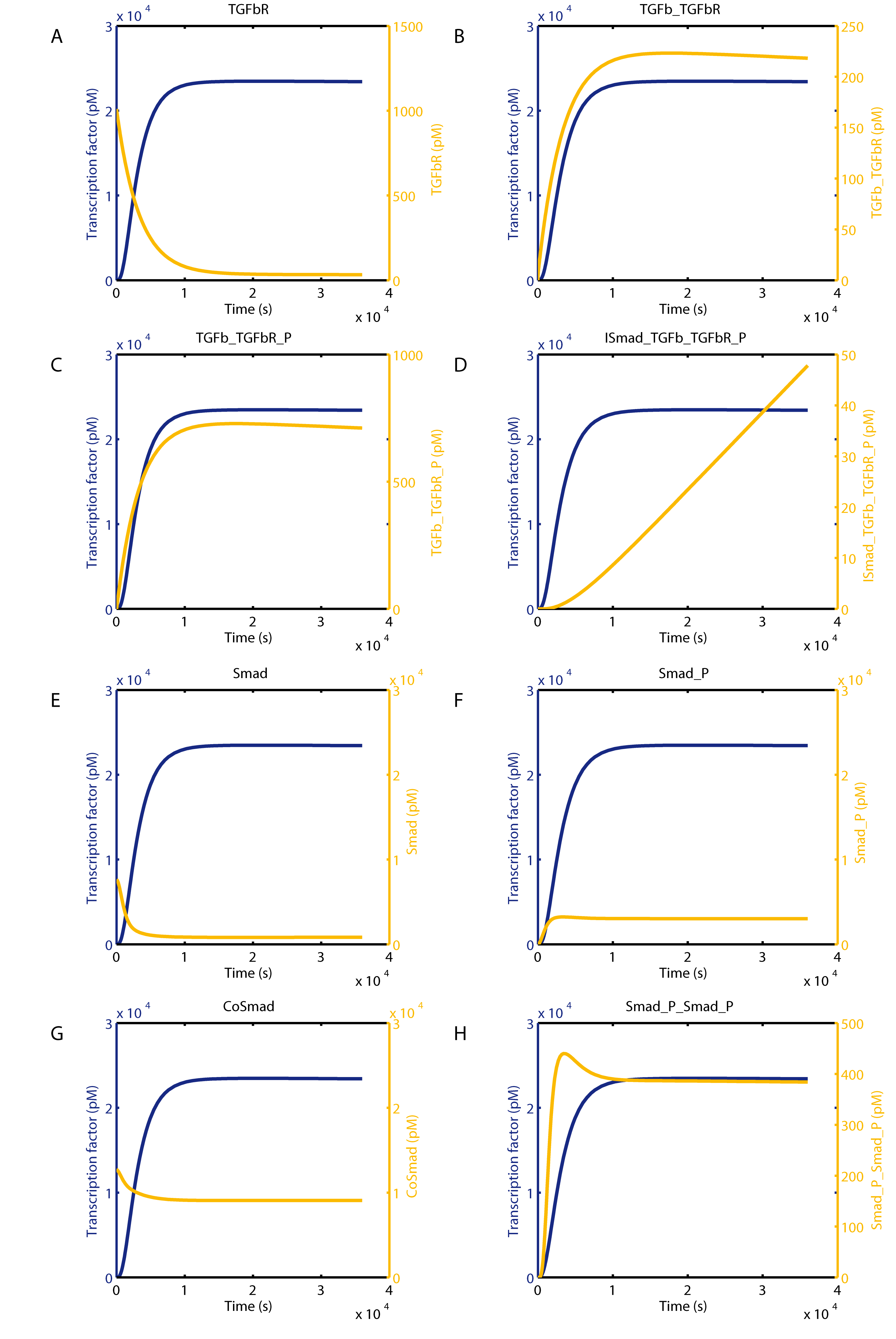}
\end{center}
\caption{
{\bf Evolution of the concentrations of all species in a representative sustained response case.} Note that different scales are chosen for each graph. The parameter set used is listed in Additional file 1, Table S3.
}
\label{supp11}
\end{figure}

\begin{figure}[!ht]
\begin{center}
\includegraphics[width=13cm]{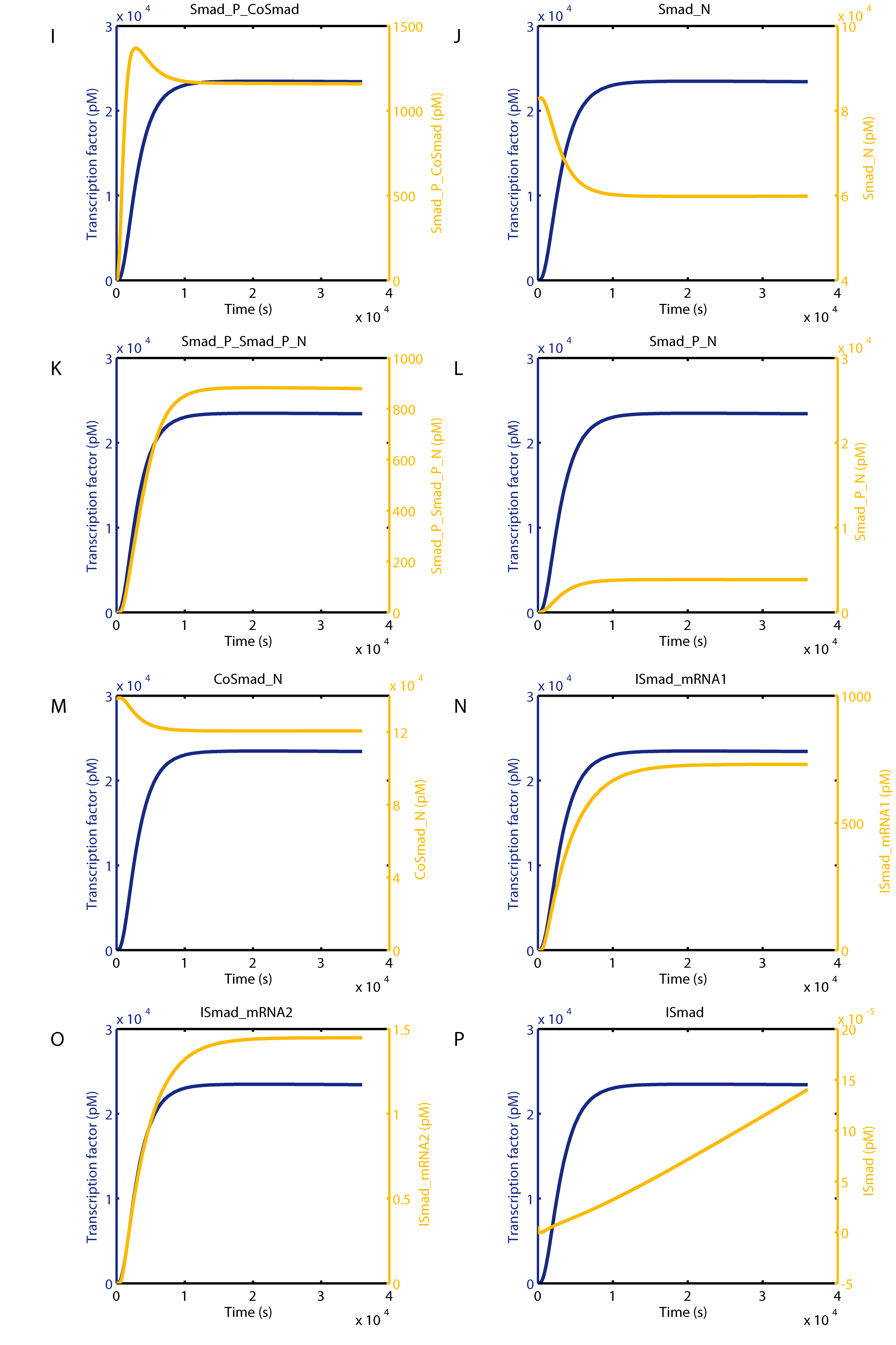}
\end{center}
\caption{
{\bf Evolution of the concentrations of all species in a representative sustained response case (continued).} Note that different scales are chosen for each graph. The parameter set used is listed in Additional file 1, Table S3.
}
\label{supp12}
\end{figure}

\begin{figure}[!ht]
\begin{center}
\includegraphics[width=15cm]{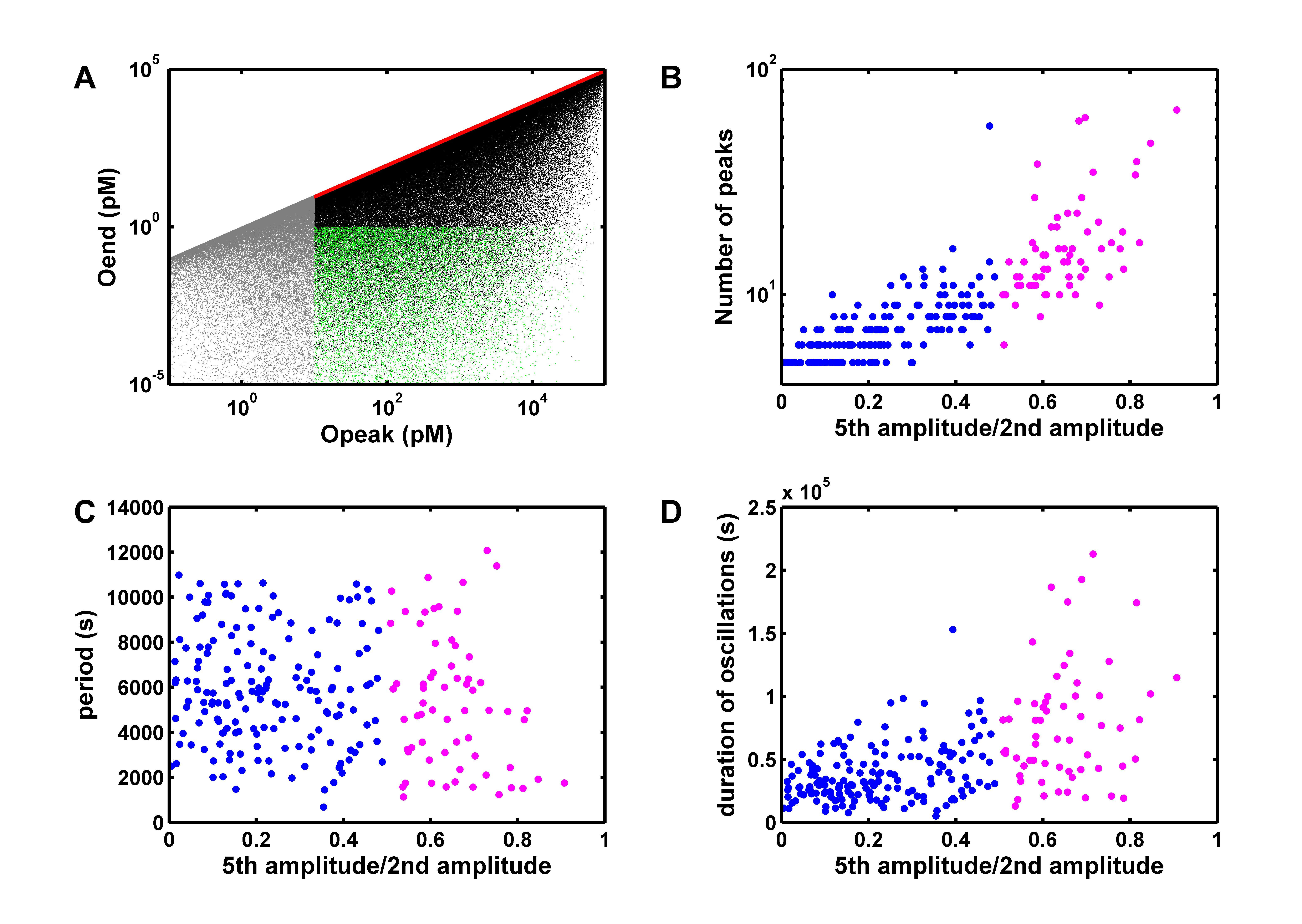}
\end{center}
\caption{
{\bf Response classification of the Parameter screen.} Based on the definitions in Fig.~2 the parameter sets can be classified to give rise to \textbf{(A)}  sustained (red), transient (green), or no responses (grey) [black parameter sets are undefined according to these criteria], or to \textbf{(B)} dampened (blue) or sustained (magenta) oscillations. Dampened oscillations (blue) have lower number of peaks before they completely vanish, compared to more sustained oscillations (purple). \textbf{(C)} Sustained (purple) and dampened (blue) oscillations exhibit the same range of periods, and \textbf{(D)} the time at which oscillation vanishes (duration) is slightly smaller for dampened oscillations. 
}
\label{Fig3}
\end{figure}

    \begin{figure}[!ht]
\begin{center}
\includegraphics[width=15cm]{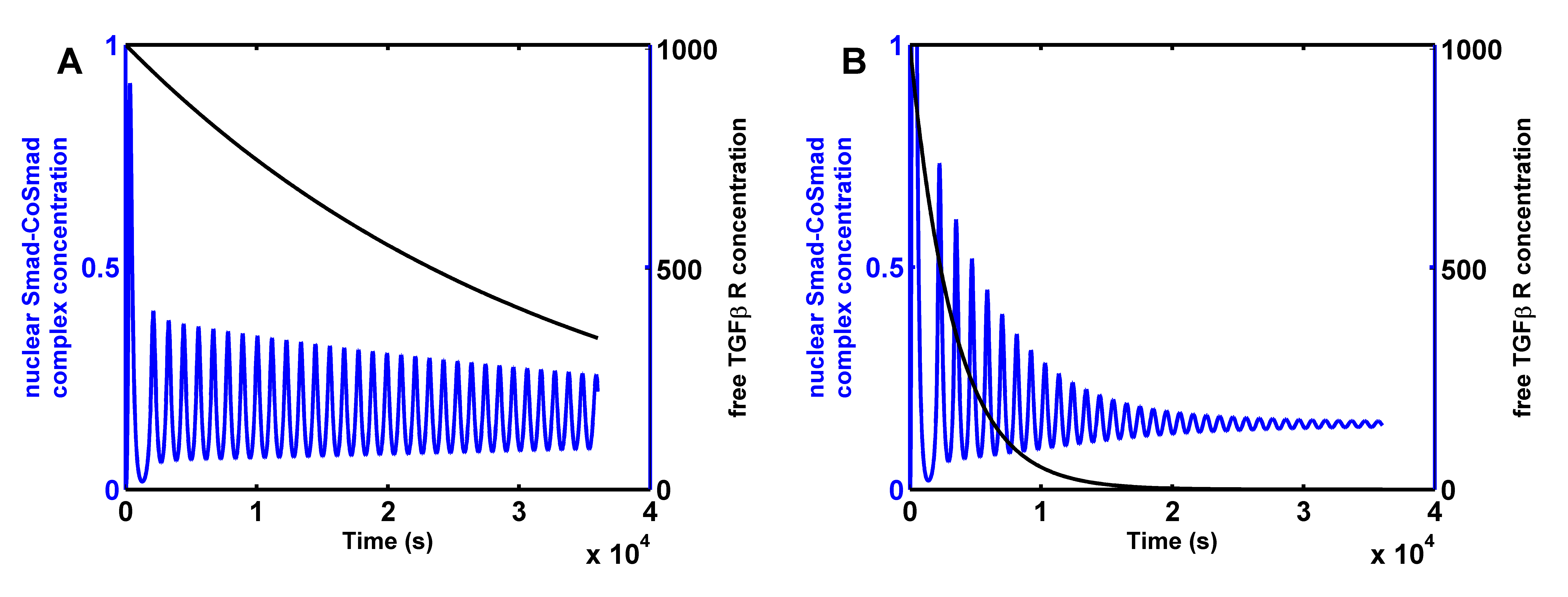}
\end{center}
\caption{
{\bf Damping of oscillations mainly depend on $k2$, the binding rate of TGF-$\beta$ to its receptor.} \textbf{(A)} Transcription factor concentration (blue) and free $TGF\beta R$ (black) evolution over time when $k2$ is small. Oscillations are sustained and the pool of free receptor decreases slowly. \textbf{(B)} Transcription factor concentration (blue) and free $TGF\beta R$ (black) evolution over time when $k2$ is 10 times larger than in A. Oscillations are dampened because quickly there is no free receptor available to bind to the ligand to induce a new peak in the transcription factor nuclear concentration.
}
\label{supp3}
\end{figure}

    \begin{figure}[!ht]
\begin{center}
\includegraphics[width=11cm]{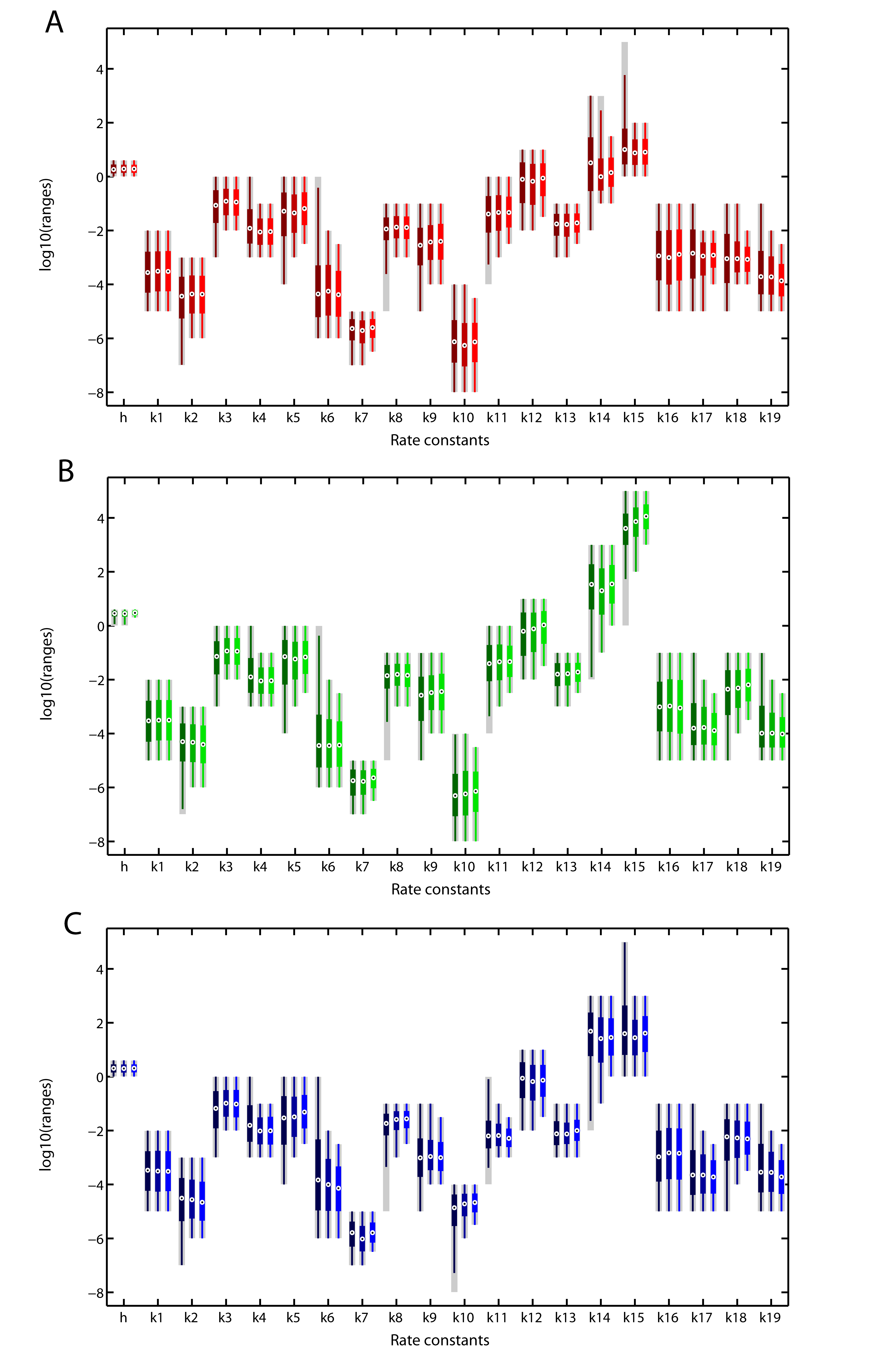}
\end{center}
\caption{
{\bf Parameter sets that can lead to both transient and sustained responses are constrained.} Box plots of parameter sets that can lead to both transient and sustained response when \textbf{(A)} $TGF\beta R$, \textbf{(B)} R-Smad, or \textbf{(C)} Co-Smad concentration is 100 fold higher or lower compared to the value of the initial screening. From the first round (dark colors) to the third round (light colors) the screening ranges are more and more reduced, depending on the boxplot of the previous round, in order to find parameter ranges that maximize the number of parameter sets switching. Ranges of the uniform sampling distributions are indicated by grey boxes.}
\label{supp4}
\end{figure}

\begin{figure}[!ht]
\begin{center}
\includegraphics[width=15cm]{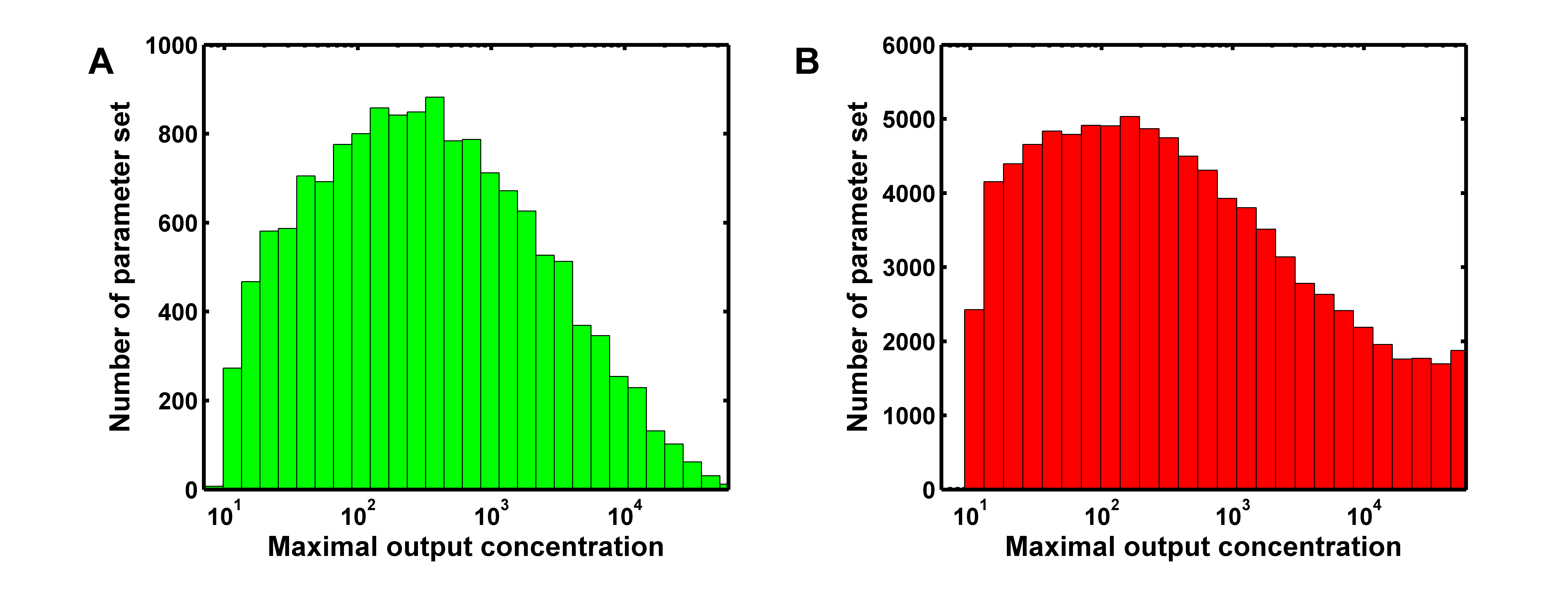}
\end{center}
\caption{
{\bf Distribution of the maximal output value for the transient (A) and sustained (B) ensemble.} Different TGF-$\beta$ concentration were applied to the system and the maximal value that the output could reach was determined. For more details see Fig.~6.\textbf{(A)} The distribution of the maximal output value for the transient ensemble is shown. For most of the sets, this value is comprised between 100 and 1000pM. \textbf{(B)} The distribution of the maximal output value for the sustained ensemble is shown. For most of the sets, this value is around 100pM, a bit lower than for the transient set.
}
\label{supp5}
\end{figure}

\begin{figure}[!ht]
\begin{center}
\includegraphics[width=15cm]{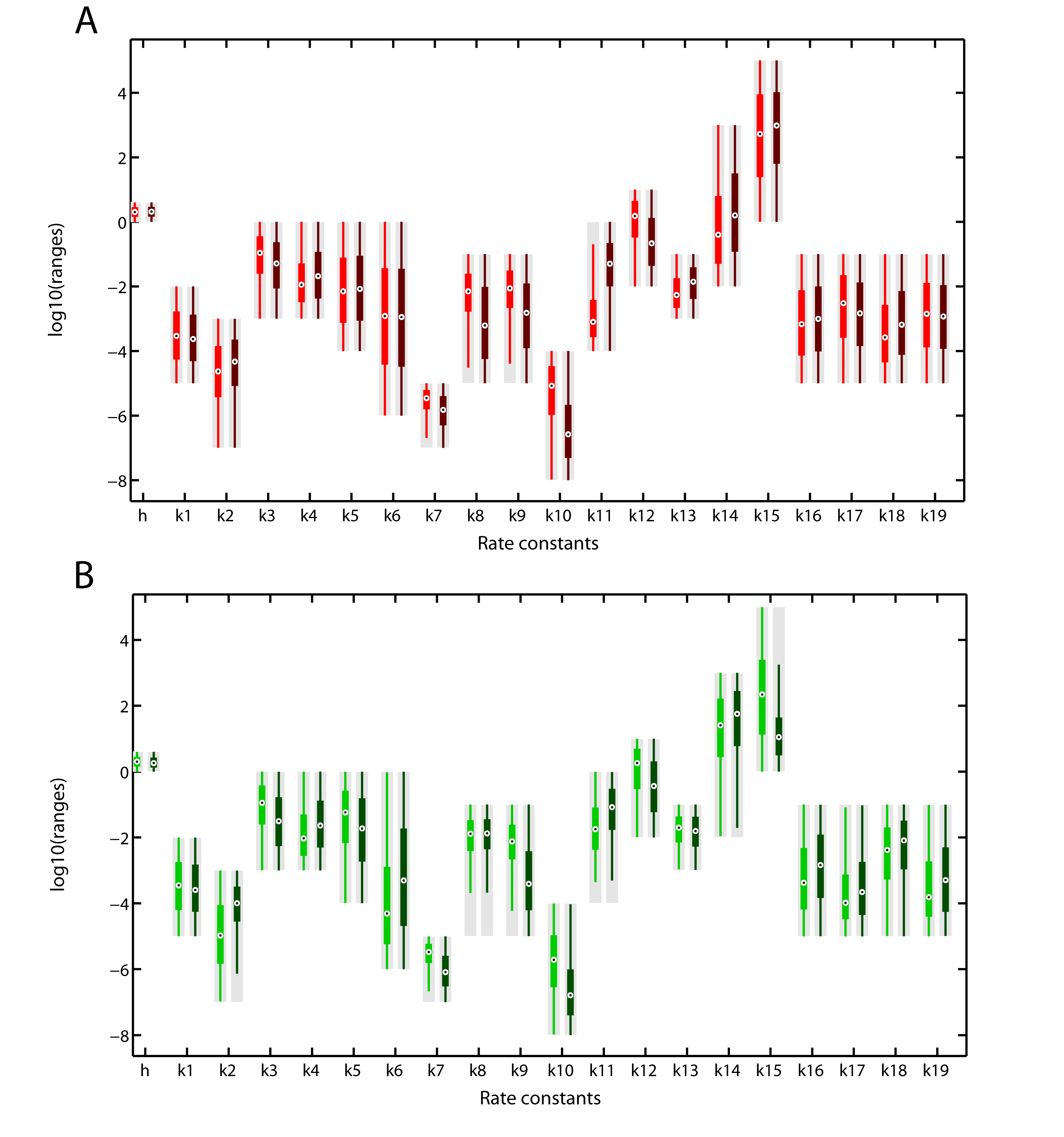}
\end{center}
\caption{
{\bf Box plots of (A) sustained and (B) transient parameter sets with high (light colors) or low (dark colors) maximal reachable output value.} Different TGF-$\beta$ concentration were applied to the system and the maximal value that the output could reach was determined. For more details see Fig.~6. \textbf{(A)} For the parameter set classified as sustained in the initial screening (Additional file 2, Fig.~S5), sets with high (light red) or low (dark red) maximal output value are boxplotted. Parameters that differ are mainly $k10$-$k11$ (formation and dissociation of the Smad/Smad and Smad/CoSmad complexes), and $k8$-$k9$ (shuttling rate from cytoplasm into the nucleus and vice-versa). \textbf{(B)} For the parameter set classified as transient in the initial screening (Additional file 2, Fig.~S5), sets with high (light green) or low (dark green) maximal output value are boxplotted. Parameters that differ are mainly $k2$ (binding of TGF-$\beta$ to its receptor), $k9$ (shuttling into the cytoplasm), $k10$ (formation of the Smad/Smad and Smad/CoSmad complexes) and $k15$ (I-Smad mRNA synthesis). Ranges of the uniform sampling distributions are indicated by grey boxes. }
\label{supp7}
\end{figure}

\begin{figure}[!ht]
\begin{center}
\includegraphics[width=15cm]{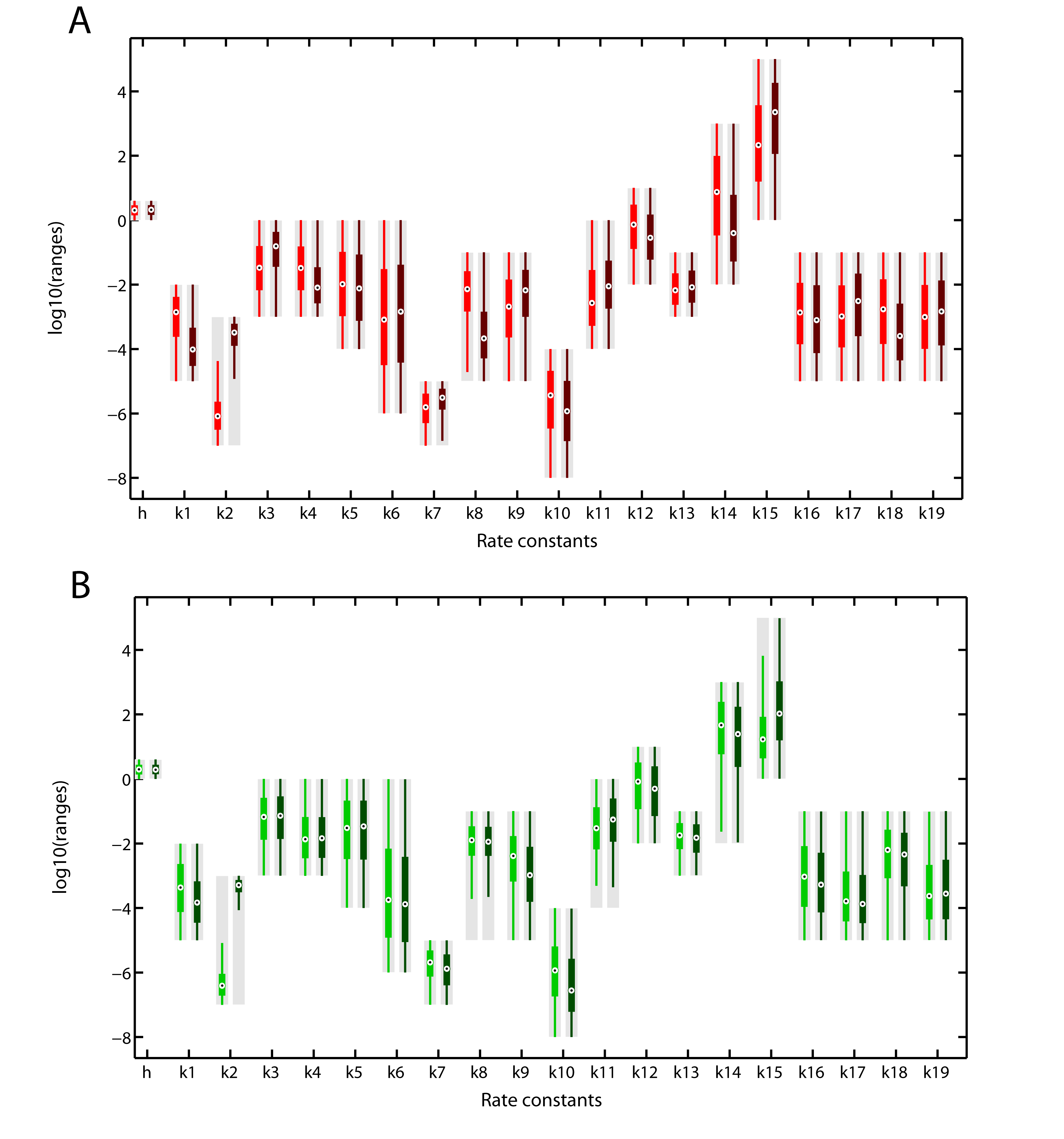}
\end{center}
\caption{
{\bf Box plots of (A) sustained and (B) transient parameter sets with high (light colors) or low (dark colors) TGF-$\beta$ saturation concentration.} Different TGF-$\beta$ concentration were applied to the system and the ligand concentration above which the output peak value cannot increase anymore was determined. For more details see Fig.~6. \textbf{(A)} For the parameter set classified as sustained in the initial screening (Additional file 2, Fig.~S5), sets with high (light red) or low (dark red) TGF-$\beta$ saturation concentration are boxplotted. Parameters that differ are mainly $k1$-$k2$ (affinity of TGF-$\beta$ for its receptor), $k3$-$k4$ (phosphorylation affinity of the receptor bound to TGF-$\beta$), and $k8$ (shuttling rate into the nucleus). \textbf{(B)} For the parameter set classified as transient in the initial screening (Additional file 2, Fig.~S5), sets with high (light green) or low (dark green) TGF-$\beta$ saturation concentration are boxplotted. The only parameter that differs significantly is $k2$ (the binding rate of TGF-$\beta$ to the receptor). Ranges of the uniform sampling distributions are indicated by grey boxes.}
\label{supp6}
\end{figure}

\begin{figure}[!ht]
\begin{center}
\includegraphics[width=15cm]{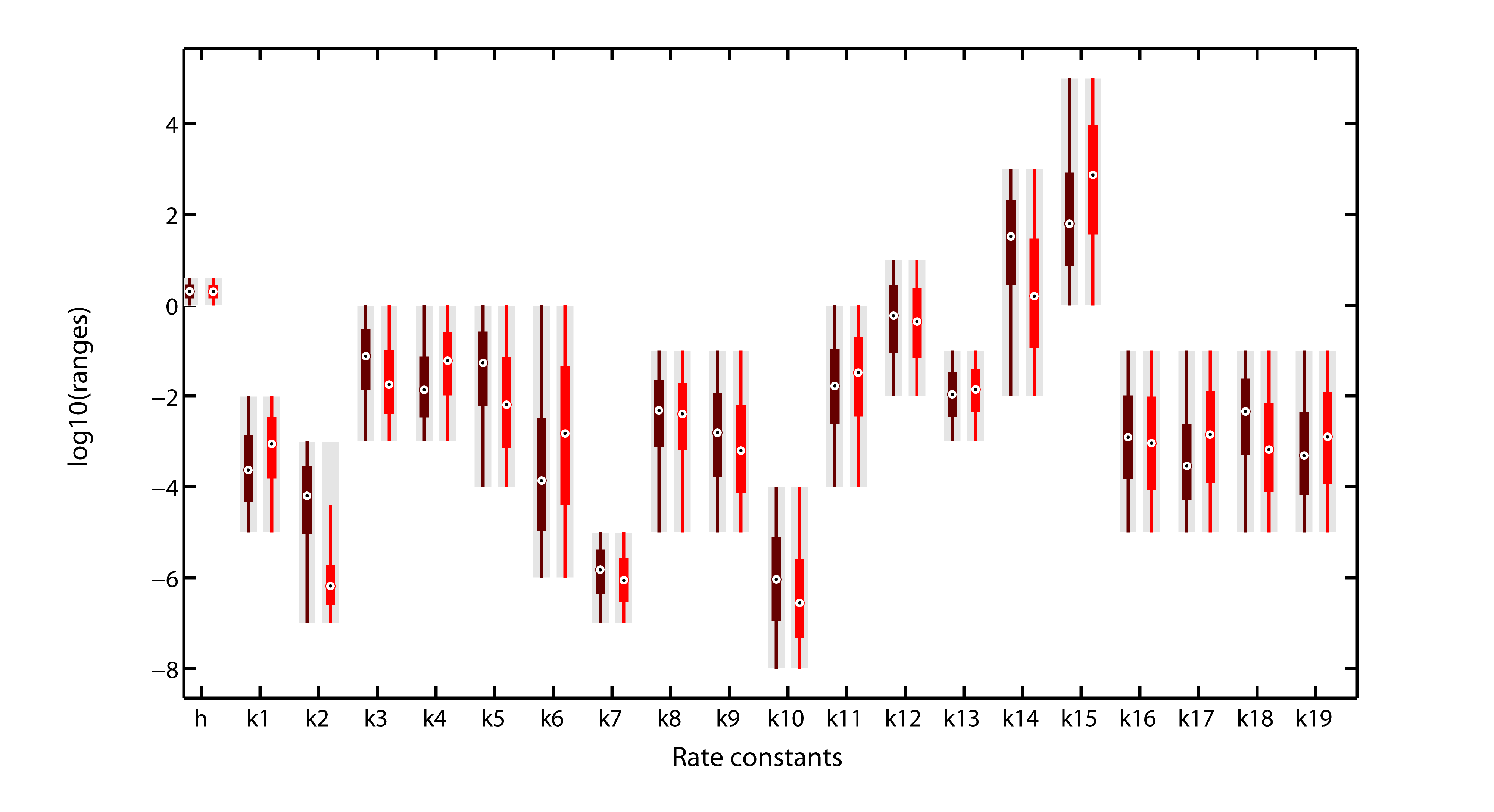}
\end{center}
\caption{
{\bf Box plots of faithful (light red) and unfaithful (dark red) parameter sets.} Procedure to classify the parameter sets into faithful or unfaithful is explained in Fig.~7A. Parameters that differ are mainly $k1$-$k2$ (affinity of TGF-$\beta$ for its receptor), $k3$-$k4$ (phosphorylation affinity of the receptor bound to TGF-$\beta$), $k5$-$k6$ (affinity of the inhibitor for the receptor bound to TGF-$\beta$) and $k15$, $k17$, $k18$, $k19$ (all related to the strenght of the feedback). Ranges of the uniform sampling distributions are indicated by grey boxes. }
\label{supp8}
\end{figure}


\renewcommand{\thefigure}{S\arabic{figure}}
\setcounter{figure}{0}


\title{Supplementary tables \\ The plasticity of TGF-$\beta$ signaling}
 

\author{Geraldine Celli\`ere, Georgios Fengos, Marianne Herv\'e and Dagmar Iber}
\date{}
\maketitle

Table S1: \textbf{Model parameters.} Parameter names, units, ranges and literature values with their references.\\[0.3cm]

Table S2: \textbf{Reaction equations used in the model.} R-Smad is denoted as Smad, $TGF\beta R$ represents the receptor, an underscore between two species indicates the complex of both species, $\_$P stands for phosphorylated proteins and $\_$N symbolize the nuclear location. When no nuclear location is specified, the name depicts the cytoplasmic species. $c = V_c / V_{ref}$ and $n = V_n / V_{ref}$ account for the volume difference between nucleus and cytoplasm as we work with concentrations. Here $V_c$ and $V_n$ refer to the cytoplasmic and nuclear volumes while $V_{ref}$ is a reference volume.
\\[0.3cm]

Table S3 : \textbf{Parameters used in Additional file 1, Fig. S8, S9, S10, and S11.} A representative transient and a representative sustained response were selected and the corresponding parameter sets were used to plot the temporal evolution of each species in both cases.

    \par
\renewcommand{\arraystretch}{1.5}
\begin{table}
\begin{tabular}{|c|c|c|c|c|c|}
\hline 
Parameters & Units & Minimum & Maximum & Literature & References\tabularnewline
\hline
\hline 
c & - & $2.3$ & $2.3$ & $2.3$ & \cite{Schmierer2008} \tabularnewline
\hline 
n & - & $1$ & $1$ & $1$ & \cite{Schmierer2008} \tabularnewline
\hline 
h &  & $1$ & $4$ & - & -\tabularnewline
\hline 
k1 & $s^{-1}$ & $10^{-5}$ & $10^{-2}$ & $2.3\times10^{-5}$ and $5\times 10^{-3}$ & \cite{Clarke2008, Chung2009} \tabularnewline
\hline 
k2 & $pM^{-1}.s^{-1}$ & $10^{-7}$ & $10^{-3}$ & $1.5\times10^{-4}$ and $1.54\times 10^{-4}$ & \cite{Clarke2008, Chung2009} \tabularnewline
\hline 
k3 & $s^{-1}$ & $10^{-3}$ & $1$ & - & -\tabularnewline
\hline 
k4 & $s^{-1}$ & $10^{-3}$ & $1$ & - & -\tabularnewline
\hline 
k5 & $pM.s^{-1}$ & $10^{-4}$ & $1$ & - & -\tabularnewline
\hline 
k6 & $s^{-1}$ & $10^{-6}$ & $1$ & - & -\tabularnewline
\hline 
k7 & $pM.s^{-1}$ & $10^{-7}$ & $10^{-5}$ & $4\times10^{-7}$ and $3.5\times 10^{-6}$ & \cite{Schmierer2008, Chung2009} \tabularnewline
\hline 
k8 & $s^{-1}$ & $10^{-5}$ & $10^{-1}$ & $2.6\times10^{-3}$and $8.3\times10^{-5}$ and $2.7\times 10^{-3}$ & \cite{Schmierer2008, Clarke2006, Chung2009}\tabularnewline
\hline 
k9 & $s^{-1}$ & $10^{-5}$ & $10^{-1}$ & $5.6\times10^{-3}$and $9.4\times10^{-2}$ and $5.8\times 10^{-3}$ & \cite{Schmierer2008, Clarke2006, Chung2009} \tabularnewline
\hline 
k10 & $pM^{-1}.s^{-1}$ & $10^{-8}$ & $10^{-4}$ & $1.6\times10^{-5}$and $1.4\times10^{-6}$ and $3.9\times 10^{-6}$ & \cite{Schmierer2008, Clarke2006, Chung2009}\tabularnewline
\hline 
k11 & $s^{-1}$ & $10^{-4}$ & $1$ & $1.8\times10^{-6}$and $7.5\times10^{-4}$ and $1.5\times 10^{-2}$  & \cite{Schmierer2008, Clarke2006, Chung2009}\tabularnewline
\hline 
k12 & - & $10^{-2}$ & $10$ & $5.7$ & \cite{Schmierer2008} \tabularnewline
\hline 
k13 & $s^{-1}$ & $10^{-3}$ & $10^{-1}$ & $4.2\times 10^{-4}$ & \cite{Chung2009}\tabularnewline
\hline 
k14 & $pM.s^{-1}$ & $10^{-2}$ & $10^{3}$ & - & -\tabularnewline
\hline 
k15 & $pM$ & $1$ & $10^{5}$ & - & -\tabularnewline
\hline 
k16 & $s^{-1}$ & $10^{-5}$ & $10^{-1}$ & - & -\tabularnewline
\hline 
k17 & $s^{-1}$ & $10^{-5}$ & $10^{-1}$ & - & -\tabularnewline
\hline 
k18 & $s^{-1}$ & $10^{-5}$ & $10^{-1}$ & - & -\tabularnewline
\hline 
k19 & $s^{-1}$ & $10^{-5}$ & $10^{-1}$ & - & -\tabularnewline
\hline 
$TGF\beta R$ & $pM$ & $10^{3}$ & $10^{3}$ & $10^{3}$, $4\times10^{3}$and $10^{4}$ & \cite{Schmierer2008, Clarke2006, Clarke2008}\tabularnewline
\hline 
Smad & $pM$ & $6\times10^{4}$ & $6\times10^{4}$ & $1.78\times10^{5}$, $3.6\times10^{5}$, $10^{5}$and $1.5\times10^{5}$ & \cite{Schmierer2008, Zi2007, Clarke2008, Clarke2006} \tabularnewline
\hline 
Cosmad & $pM$ & $10^{5}$ & $10^{5}$ & $10^{5}$, $8.4\times10^{5}$, $10^{5}$and $1.5\times10^{5}$ & \cite{Schmierer2008, Zi2007, Clarke2006, Clarke2008} \tabularnewline
\hline 
TGF-$\beta$ & $pM$ & $200$ & $200$ & $80$ & \cite{Zi2007} \tabularnewline
\hline
\end{tabular} 
\caption{{\bf Model parameters.}}
\label{Tab1}
\end{table}


\begin{landscape}

\begin{table}
\renewcommand{\arraystretch}{2}
\small
\begin{tabular}{|l|ll|}
\hline 
\multicolumn{1}{|c|}{Differential equations} &  \multicolumn{2}{|c|}{Reactions definitions}\tabularnewline
\hline
$\dfrac{d[TGF\beta R]}{dt}=r1-r2$ & $r1=k1\times [TGF\beta\_TGF\beta R]$ & $r18=k9\times [Smad\_P\_N]$ \tabularnewline      
$\dfrac{d[TGF\beta\_TGF\beta R]}{dt}=-r1+r2-r3+r4+r6$  & $r2=k2\times [TGF\beta R]\times [TGF\beta]$ & $r19=k12\times k8\times [Smad\_P\_CoSmad]$\tabularnewline
$\dfrac{d[TGF\beta\_TGF\beta R\_P]}{dt}=r3-r4-r5$  & $r3=k3\times [TGF\beta\_TGF\beta R]$ & $r20=k13\times [Smad\_P\_N]$\tabularnewline
$\dfrac{d[I\_Smad\_TGF\beta\_TGF\beta R\_P]}{dt}=r5-r6$   & $r4=k4\times [TGF\beta\_TGF\beta R\_P]$ & $r21=k10\times [Smad\_P\_N]\times [Smad\_P\_N]$\tabularnewline                
$\dfrac{d[Smad]}{dt}=-r7-r8/c+r9/c$  & $r5=k5\times [TGF\beta\_TGF\beta R\_P]\times [I\_Smad]$ & $r22=k11\times [Smad\_P\_Smad\_P\_N]$\tabularnewline             
$\dfrac{d[Smad\_P]}{dt}=r7-r10+r11-r12+r13-r17/c+r18/c$ & $r6=k6\times [I\_Smad\_TGF\beta\_TGF\beta R\_P]$ & $r23=k10\times [Smad\_P\_N]\times [CoSmad\_N]$\tabularnewline                        
$\dfrac{d[CoSmad]}{dt}=-r12+r13-r14/c+r15/c$  & $r7=k7\times [Smad]\times [TGF\beta\_TGF\beta R\_P]$ & $r24=k11\times [Smad\_P\_CoSmad\_N]$\tabularnewline   
$\dfrac{d[Smad\_P\_Smad\_P]}{dt}=r10-r11-r16/c$ & $r8=k8\times [Smad]$ & $r25=k14\times \dfrac{[Smad\_P\_CoSmad\_N]^{h}}{[Smad\_P\_CoSmad\_N]^{h}+k15^{h}}$\tabularnewline               
$\dfrac{d[Smad\_P\_CoSmad]}{dt}=r12-r13-r19/c$ & $r9=k9\times [Smad\_N]$ & $r26=k16\times [I\_Smad\_mRNA1]$\tabularnewline               
$\dfrac{d[Smad\_N]}{dt}=r8/n-r9/n+r20$ & $r10=k10\times [Smad\_P]\times [Smad\_P]$ & $r27=k17\times [I\_Smad\_mRNA2]$\tabularnewline               
$\dfrac{d[Smad\_P\_Smad\_P\_N]}{dt}=r16/n+r21-r22$ & $r11=k11\times [Smad\_P\_Smad\_P]$ & $r28=k18\times [I\_Smad\_mRNA2]$\tabularnewline                      
$\dfrac{d[Smad\_P\_N]}{dt}=r17/n-r18/n-r20-r21+r22-r23+r24$ & $r12=k10\times [Smad\_P]\times [CoSmad]$ & $r29=k19\times [I\_Smad]$\tabularnewline
$\dfrac{d[Smad\_P\_CoSmad\_N]}{dt}=r19/n+r23-r24$ & $r13=k11\times [Smad\_P\_CoSmad]$ & \tabularnewline
$\dfrac{d[CoSmad\_N]}{dt}=r14/n-r15/n-r23+r24$ & $r14=k8\times [CoSmad]$ & \tabularnewline
$\dfrac{d[I\_Smad\_mRNA1]}{dt}=r25-r26/n$ & $r15=k9\times [CoSmad\_N]$ & \tabularnewline             
$\dfrac{d[I\_Smad\_mRNA2]}{dt}=r26/c-r27$ & $r16=k12\times k8\times [Smad\_P\_Smad\_P]$ & \tabularnewline                   
$\dfrac{d[I\_Smad]}{dt}=r28-r29+r6-r5$ & $r17=k8\times [Smad\_P]$      & \tabularnewline                   
\hline            
\end{tabular}
\caption{ {\bf Reaction equations used in the model.}}
\label{Tab2}
\end{table}

\end{landscape}
\newpage

\normalsize
\par
\begin{table}
\begin{center}
\begin{tabular}{|c|c|c|}
\hline 
Parameters & Transient Response & Sustained Response\tabularnewline
\hline
\hline 
h & $2.06$ & $1.32$\tabularnewline
\hline 
k1 & $4.46\times10^{-3}$ & $4.41\times10^{-5}$\tabularnewline
\hline 
k2 & $4.39\times10^{-6}$ & $1.47\times10^{-6}$\tabularnewline
\hline 
k3 & $3.24\times10^{-1}$ & $3.62\times10^{-2}$\tabularnewline
\hline 
k4 & $1.92\times10^{-3}$ & $1.11\times10^{-2}$\tabularnewline
\hline 
k5 & $5.49\times10^{-4}$ & $2.40\times10^{-1}$\tabularnewline
\hline 
k6 & $1.29\times10^{-5}$ & $4.69\times10^{-4}$\tabularnewline
\hline 
k7 & $9.35\times10^{-6}$ & $6.44\times10^{-6}$\tabularnewline
\hline 
k8 & $1.04\times10^{-2}$ & $2.05\times10^{-3}$\tabularnewline
\hline 
k9 & $7.50\times10^{-4}$ & $1.74\times10^{-4}$\tabularnewline
\hline 
k10 & $5.12\times10^{-8}$ & $2.77\times10^{-7}$\tabularnewline
\hline 
k11 & $9.23\times10^{-3}$ & $5.61\times10^{-3}$\tabularnewline
\hline 
k12 & $5.13\times10^{-2}$ & $1.02$\tabularnewline
\hline 
k13 & $1.64\times10^{-3}$ & $2.26\times10^{-3}$\tabularnewline
\hline 
k14 & $3.80\times10^{-2}$ & $2.04\times10^{-1}$\tabularnewline
\hline 
k15 & $28.52$ & $1131.8$\tabularnewline
\hline 
k16 & $2.14\times10^{-2}$ & $2.74\times10^{-4}$\tabularnewline
\hline 
k17 & $8.05\times10^{-5}$ & $6.02\times10^{-2}$\tabularnewline
\hline 
k18 & $4.34\times10^{-2}$ & $1.05\times10^{-3}$\tabularnewline
\hline 
k19 & $4.12\times10^{-4}$ & $1.21\times10^{-5}$\tabularnewline
\hline
\end{tabular} 

\caption{\textbf{Parameters used in Additional file 1, Fig. S8, S9, S10, and S11.} 
}
\end{center}
\end{table}

\clearpage
{\ifthenelse{\boolean{publ}}{\footnotesize}{\small}
 \bibliographystyle{bmc_article}  
\bibliography{./biblio}